\documentclass[lettersize,journal]{IEEEtran}
\usepackage{amsmath,amsfonts}
\usepackage{algorithmic}
\usepackage{algorithm}
\usepackage{array}
\usepackage[caption=false,font=normalsize,labelfont=sf,textfont=sf]{subfig}
\usepackage{textcomp}
\usepackage{stfloats}
\usepackage{url}
\usepackage{verbatim}
\usepackage{graphicx}
\usepackage{cite}
\usepackage{booktabs}
\usepackage{svg}
\usepackage{makecell}
\usepackage{amssymb}
\usepackage{threeparttable}
\usepackage{multirow}
\usepackage[colorlinks=true, citecolor=blue, linkcolor=blue, urlcolor=black, pdfborder={0 0 0}]{hyperref}
\definecolor{changecolor}{RGB}{255, 0, 0}
\newcommand{\change}[1]{#1}
\begin{document}
	
	\title{Global Rotation Equivariant Phase Modeling for Speech Enhancement with Deep Magnitude-Phase Interaction}
	
	\author{Chengzhong Wang, Andong Li, {\IEEEmembership{Member, IEEE}}, Dingding Yao, and Junfeng Li
		\thanks{This research is partially supported by National Natural Science Foundation of China (No. 12425413, 12574509), Beijing Natural Science Foundation (L257001, L247022), Oriented Project Independently Deployed by Institute of Acoustics, Chinese Academy of Sciences (MBDX202402).}
		\thanks{Chengzhong Wang, Dingding Yao and Junfeng Li are
			with the Key Laboratory of Speech Acoustics and Content Understanding, Institute of Acoustics, Chinese Academy of Sciences, Beijing 100190, China, and also with University of Chinese Academy of Sciences, Beijing 100049, China. (Email: wangchengzhong@hccl.ioa.ac.cn, yaodingding@hccl.ioa.ac.cn, lijunfeng@hccl.ioa.ac.cn)}
		\thanks{Andong Li is
			with the Key Laboratory of Noise and Vibration Research, Institute of Acoustics, Chinese Academy of Sciences, Beijing 100190, China, and also with University of Chinese Academy of Sciences, Beijing 100049, China. (Email: liandong@mail.ioa.ac.cn)}
		\thanks{Corresponding Author: Junfeng Li.}}
	
	\markboth{}%
	{Shell \MakeLowercase{\textit{et al.}}: A Sample Article Using IEEEtran.cls for IEEE Journals}
	
	
	\maketitle
	
	\begin{abstract}

		While deep learning has advanced speech enhancement (SE), effective phase modeling remains challenging, as conventional networks typically operate within a flat Euclidean feature space, which is not easy to model the underlying circular topology of the phase. To address this, we propose \change{a magnitude-phase dual-stream framework} that aligns the phase stream with its intrinsic circular geometry by enforcing Global Rotation Equivariance (GRE) characteristic. Specifically, we introduce a Magnitude-Phase Interactive Convolutional Module (MPICM) for modulus-based information exchange and a Hybrid-Attention Dual \change{Feed-Forward Network} (HADF) bottleneck for unified feature fusion, both of which are designed to preserve  GRE in the phase stream. Comprehensive evaluations are conducted across phase retrieval, denoising, dereverberation, and bandwidth extension tasks to validate the superiority of the proposed method over multiple advanced baselines. Notably, the proposed architecture reduces Phase Distance by over 20\% in the phase retrieval task and improves PESQ by more than 0.1 in zero-shot cross-corpus denoising evaluations. The overall superiority is also established in universal SE tasks involving mixed distortions. Qualitative analysis further reveals that the learned phase features exhibit distinct periodic patterns, which are consistent with the intrinsic circular nature of the phase. The source code is available at \url{https://github.com/wangchengzhong/GRE-Net}.
	\end{abstract}
	
	\begin{IEEEkeywords}
		Speech enhancement, phase modeling, global rotation equivariance, magnitude-phase interaction, complex-valued neural network.
	\end{IEEEkeywords}
	
	\section{Introduction}
	\IEEEPARstart{s}{peech} enhancement (SE), the process of improving the quality and intelligibility of speech degraded by acoustic distortion, is a critical technology for telecommunications, smart devices, and hearing aids. SE has historically relied on traditional signal processing methods \cite{ephraim1984speech, cohen2001speech}. However, these conventional approaches often struggle in highly non-stationary acoustic environments. The advent of deep learning has precipitated a paradigm shift, redefining SE as a data-driven discipline. State-of-the-art methods typically learn non-linear masking \cite{Wang2014training} or mapping \cite{xu2014regression} functions from degraded to clean speech using vast datasets of paired examples. These data-driven approaches demonstrate superior performance over traditional techniques, particularly in low signal-to-noise ratio conditions and non-stationary degradation \cite{zheng2023sixty}.
	
	These methods can be broadly categorized into the time and time-frequency domains. Approaches in the T-F domain are often favored for their physical interpretability \cite{li2021two} and robustness in noisy-reverberant scenarios\cite{maciejewski2020whamr}. Historically, T-F domain methods prioritized magnitude processing \cite{zheng2023sixty}, disregarding phase due to modeling complexities and its assumed insignificance relative to magnitude \cite{wang1982the}. However, with a growing understanding of the role phase plays in speech perception, its importance in challenging reverberant and low signal-to-noise ratio (SNR) conditions has become evident \cite{paliwal2011importance}. This necessity is further amplified in Universal Speech Enhancement (USE), which encompasses tasks beyond simple additive noise reduction. For instance, in dereverberation,  while magnitude-only processing can partially mitigate amplitude-related effects through appropriate compression, it cannot adequately recover the phase information needed to structurally correct these distortions \cite{li2021on}. Similarly, in bandwidth extension, the realistic regeneration of high-frequency components relies heavily on phase continuity to align harmonics and avoid perceptual artifacts \cite{lu2025towards}. Consequently, phase-aware methods have increasingly been explored \cite{gerkmann2015phase,yin2020phasen,lu2025towards}, achieving better performance by fully exploiting the phase information inherent in speech signals.
	
	Tracing the evolution of state-of-the-art predictive SE methods reveals a clear progression toward disentangled magnitude and phase modeling. The Complex Two-Stage Network (CTS-Net) \cite{li2021two} pioneered the idea of magnitude estimation and complex refinement, while the Dual-Branch Attention-In-Attention Transformer (DB-AIAT) \cite{yu2022dual} introduced a parallel decoder paradigm, which was subsequently adopted by works such as CMGAN \cite{abdulatif2024cmgan}. Later, parallel Magnitude-Phase estimation Network (MP-SENet) \cite{lu2023mp} advanced this strategy by explicit phase modeling: it excluded magnitude information from the complex branch to create a dedicated phase decoder regulated by phase-specific loss functions \cite{ai2023neural}. Despite sharing a similar architectural backbone, MP-SENet achieved significant performance gains from the explicit phase modeling approach. 
	Following MP-SENet and its updated iteration  \cite{lu2025explicit}, recent works have made significant strides in improving computational efficiency \cite{wang2025mamba}, enhancing structural flow between the encoder and decoders \cite{wang2025zipenhancer}, and incorporating auxiliary priors \cite{hu2025mn,xu2025interactive}. While these contributions have substantially pushed the boundaries of model efficiency and representation capability, the fundamental challenge of intrinsic phase  modeling remains relatively less explored.
	
	Specifically, although current decoupled architectures split magnitude and phase processing, either via a shared bottleneck \cite{lu2025explicit} or parallel streams \cite{lu2025towards, yin2020phasen}, they limit the handling of phase's distinct structural nature to the loss term \cite{ai2023neural}, treating it merely as a generic data distribution within the network itself. Mathematically, phase resides on a circular manifold, characterized by periodic continuity. In contrast, the fundamental building blocks of commonly used deep networks operate under Euclidean assumptions. Although complex-valued networks \cite{hu2020dccrn} avoid the numerical singularity of phase wrapping by predicting Cartesian components, they introduce a subtler geometric mismatch. Standard operations (e.g., biased convolutions) learn a ``preferred'' orientation in the complex plane even when there is no input. However, the intrinsic geometry of the phase should be insensitive to the global starting position. This topological perspective aligns with the signal processing properties of the analytic speech signal, where applying a global phase rotation alters the waveform shape  but preserves the temporal envelope and instantaneous frequency, with indistinguishable perceptual difference \cite{thien2023inter, zhang2024unrestricted}. Consequently, extensive literature models phase primarily through its relative structure \cite{masuyama2020, ghiglia1998}. By enforcing equivariance, we ensure that the network separates relative phase structure from the absolute phase orientation. 
	
	While Zhang et al. \cite{zhang2024unrestricted} previously identified the issue of global phase bias and addressed it via data augmentation to encourage robustness, our approach offers a fundamental improvement.  Instead of using data augmentation to make the network insensitive to the coordinate system, we redesign the network inner structure to respect this topological symmetry, treating the phase space as a rotationally symmetric surface via global rotation equivariant (GRE) network design. As a result, the model becomes intrinsically insensitive to absolute phase orientation, eliminating the need to learn this equivariance from data. This design effectively creates a processing stream that treats absolute phase orientation as arbitrary while rigorously concentrating on relative structures. 
	
	Given that the rotational symmetry of phase is fundamentally distinct from the geometry of magnitude, we propose a Deep Magnitude-Phase Interaction scheme which maintains a dual-stream flow to respect these topological differences while facilitating necessary information exchange between them. We introduce two novel components: the Magnitude-Phase Interactive Convolutional Module (MPICM) and the Hybrid-Attention Dual \change{Feed-Forward Network} (HADF) structure. These modules allow for the rigorous exchange of information without breaking the rotation-equivariant symmetry of the phase stream. Motivated by recent trends in universal speech restoration \cite{liu2025universal}, we evaluate our proposed model across a comprehensive suite of tasks that involve phase recovery with complete or distorted magnitude: phase retrieval, denoising, dereverberation, bandwidth extension, and mixed distortions. The main contributions of this paper are summarized as follows:
	\begin{itemize}
		\item We propose a phase processing framework \change{for speech enhancement} that explicitly respects the circular topology of phase by enforcing Global Rotation Equivariance as a structural inductive bias throughout the phase stream. This design aligns the network with the intrinsic circular geometry of the phase, effectively \change{eliminating} the topological mismatch inherent in conventional Euclidean deep networks.
		\item We devise a Deep Magnitude-Phase Interaction scheme comprising the Magnitude-Phase Interactive Convolutional Module (MPICM) and the Hybrid-Attention Dual \change{Feed-Forward Network} (HADF). These modules enable rigorous interactive information exchange via cross-branch modulus gating and unified attention score domain interaction respectively, while strictly preserving the geometric constraints of the phase stream.
		\item We demonstrate that our method achieves superior phase modeling accuracy and generalization capability with fewer parameters and comparable computational cost.  Notably, it yields up to a 20\% decrease of Phase Distance (PD) in phase retrieval and over 0.1 PESQ improvement in zero-shot cross-corpus denoising evaluations. When applied to phase-involved universal speech enhancement task, including denoising, dereverberation, bandwidth extension, and mixed distortions, the proposed equivariant phase modeling demonstrates clear overall advantages, \change{achieving state-of-the-art phase estimation accuracy while maintaining competitive perceptual performance across all evaluated conditions.}
	\end{itemize}

	\section{Global Rotation Equivariance}
	\label{section:global rotation equivariance}
	In this section, we establish the theoretical foundation of our method, defining the mathematical conditions for Global Rotation Equivariance and delimiting its scope relative to time-shift invariance.
	\subsection{Definition and Topological Basis}
	Global rotation Equivariance refers to the symmetry of a function with respect to global phase shifts \cite{huang2025holographic}. Let $\mathbf{x} \in \mathbb{C}^{C \times H \times W}$ denote a complex-valued input tensor with $C$ channels and $(H,W)$ spatial dimensions, and let $T_\theta$ be a global rotation operator such that $T_\theta(\mathbf{x}) = \mathbf{x} \cdot e^{j\theta}$ for any angle $\theta \in [0, 2\pi)$. A neural transformation $\mathcal{F}: \mathbb{C}^{(C,H,W)} \to \mathbb{C}^{(C',H',W')}$ is defined as global rotation-equivariant if and only if it commutes with $T_\theta$:
	\begin{equation}
		\mathcal{F}(T_\theta(\mathbf{x})) = T_\theta(\mathcal{F}(\mathbf{x})) \iff \mathcal{F}(\mathbf{x} \cdot e^{j\theta}) = \mathcal{F}(\mathbf{x}) \cdot e^{j\theta}.
		\label{eq:equivariance_def}
	\end{equation}
	The rationale for imposing this constraint for phase modeling lies in the topological nature of the phase. Unlike magnitude, which resides in Euclidean space, phase is inherently circular, residing on the manifold $S^1$ \change{\cite{klein2020torus}}. On this manifold, standard Euclidean metrics are ill-defined due to the rotational symmetry. By enforcing global rotation equivariance, the model aligns with this circular geometry, treating the absolute phase orientation as arbitrary while rigorously preserving the \textit{relative} phase differences such as GD and IP that encode signal structure. 
	
	It is important to clarify the physical interpretation of this rotation in the context of analytic speech signal. Preserving the real-valued nature of speech signal requires Hermitian symmetry in the frequency domain\change{, i.e., for spectrum X at frequency bin k, we have $X[-k] = X[k]^*$, with * representing conjugation}. A true global rotation $e^{j\theta}$ across strictly all frequencies would yield a complex time-domain signal. However, modern speech enhancement networks operate on the \textit{single-sided} spectrum, which effectively models the analytic signal.  In this domain, applying a global rotation $e^{j\theta}$ corresponds to a constant phase shift of the analytic representation (a fractional Hilbert transform\change{\cite{lohmann1996}}). Since this operation preserves the signal's temporal envelope and relative phase structure, the absolute phase orientation carries indistinguishable  information, \change{a finding also supported by recent perceptual studies \cite{zhang2024unrestricted}.} Consequently, enforcing global rotation equivariance is justified: it prevents the network from wasting capacity on the arbitrary alignment of the carrier, ensuring the learned representations are invariant to the coordinate system of the complex plane.
	
	\subsection{The Scope of Equivariance: Global vs. Frequency-Dependent}
	It is also critical to distinguish between \textit{Global Rotation} (a constant phase shift $e^{j\theta}$ across frequencies) and \textit{Frequency-Dependent Rotation} (the linear phase ramp $e^{-j\omega \tau}$ induced by a time shift $\tau$). While it is intuitively appealing to assume that speech phase modeling  prioritizes time-shift equivariance, we do not use this bias because of modeling capacity: Key structural characteristics of speech, such as onsets, manifest distinct patterns in the frequency-derivative of the phase (Group Delay, GD) \cite{maitra2011spectral}. However, these intrinsic group delay patterns are deeply coupled with the linear phase ramps induced by time shifts \cite{murthy2011group}. Therefore, enforcing equivariance to frequency-dependent rotations would deprive the model of the ability to modify the GD, restricting its expressive power.
	Therefore, we adopt Global Rotation Equivariance as the maximal valid symmetry that \change{aligns the model with the circular topology of phase while preserving the capacity for speech spectrum modeling}.
	
	\subsection{Atomic Global Rotation-Equivariant Operations}
	Constructing a deep GRE architecture requires identifying fundamental mathematical operations that satisfy (\ref{eq:equivariance_def}). \change{While a rich literature on rotation-equivariant networks has been developed for spatial transformations on real-valued features~\cite{liu2025rotation,villar2021scalars}, our setting involves global phase rotations in the complex domain. Therefore, we focus on two classes of atomic operations that are suitable and computationally convenient for enforcing global rotation equivariance in  phase modeling:}
	
	\textit{1) Bias-free Complex-Linear Transformations:} Standard complex-valued linear transformation (including convolution) is inherently rotation-equivariant, provided the additive bias is strictly zero. A non-zero bias term $\mathbf{b}$ breaks symmetry as $\mathbf{W}(\mathbf{x} e^{j\theta}) + \mathbf{b} \neq (\mathbf{W}\mathbf{x} + \mathbf{b}) e^{j\theta}$, where $\mathbf{x}$ is the input tensor. \change{Examples of such atomic operations include bias‑free complex‑valued convolution layers, which we will adopt in our network design for phase modeling.}
	
	\textit{2) Invariant-Modulated Transformations:} While holomorphic activations are restrictive \cite{lee2022complex}, non-linearities can be introduced by modulating the equivariant feature $\mathbf{x}$ element-wise with a rotation-invariant real-valued tensor $\mathcal{S}(\mathbf{x})$ (where $\mathcal{S}(\mathbf{x}e^{j\theta}) = \mathcal{S}(\mathbf{x})$). The transformation $\mathcal{F}(\mathbf{x}) = \mathbf{x} \odot \mathcal{S}(\mathbf{x})$ preserves overall equivariance:
	\begin{equation}
		\mathcal{F}(\mathbf{x}e^{j\theta}) = (\mathbf{x}e^{j\theta}) \odot \mathcal{S}(\mathbf{x}e^{j\theta}) = (\mathbf{x} \odot \mathcal{S}(\mathbf{x})) \cdot e^{j\theta}. \label{invariant-modulated transformations}
	\end{equation}
	This principle allows us to incorporate robust non-linear modeling capabilities, ranging from modulus-based gating mechanisms to rotation-invariant attention weights\change{, which we will detail later}, without corrupting the GRE property.

	\section{Proposed Method}

	\subsection{Network Overview}\label{network_overview}
	\begin{figure*}[t]
		\centering
		\includegraphics[width=\textwidth]{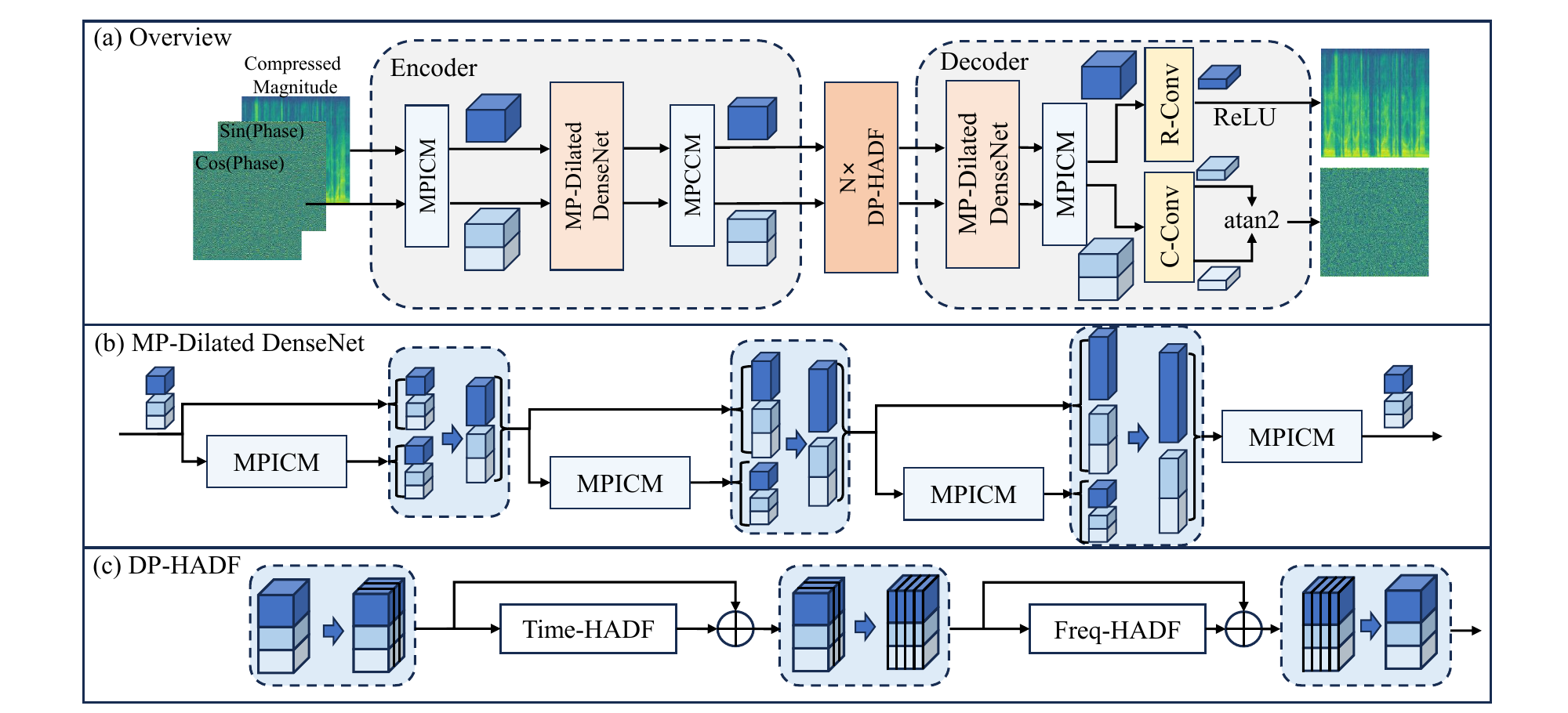}
		\caption{Overview of the proposed network architecture. (a) The dual-stream encoder-decoder topology. The R-Conv and C-Conv denote real-valued and complex-valued convolution respectively. (b) Structure of the Magnitude-Phase Dilated DenseNet, illustrating the aligned channel concatenation. (c) Signal flow within the dual-path \change{(DP)} Hybrid-Attention Dual \change{Feed-Forward Network} (HADF) bottleneck.}
		\label{fig:wang1}
	\end{figure*}
	Fig. \ref{fig:wang1} illustrates the proposed architecture, which comprises an encoder, a dual-path bottleneck, and a decoder. Unlike prior works that prioritize macro-level topological search \cite{yu2022dbt,abdulatif2024cmgan,lu2025explicit}, our contribution lies in redefining the fundamental computational basis to enforce global rotation equivariance. Magnitude and phase are treated with distinct streams maintained as real-valued and complex-valued features throughout the network, where phase stream satisfies global rotation equivariance throughout the network. It is achieved via two core innovations: 1) The Magnitude-Phase Interactive Convolutional Module (MPICM), which replaces standard convolution in the encoder and decoder; 2) the Hybrid Attention Dual \change{Feed-Forward Network} (HADF) module which serves as the dual-path bottleneck structure. 
	\change{Both modules maintain separate, interactive magnitude‑phase dual streams (detailed in Sections \ref{subsection:mpicm} and \ref{subsection:hadf}, respectively). Here we first provide a high‑level description of the overall encoder‑bottleneck‑decoder architecture built upon these modules to give an intuitive understanding before delving into the specifics.}
	
	\subsubsection{Encoder} The encoder extracts \change{time-frequency} features using the MPICM as the fundamental operator. Each MPICM takes two streams as input and output. The encoder initiates with a simplified variant of the MPICM, which processes the compressed magnitude spectrum $\mathbf{M} = \mathbf{Y}_m^{\alpha}\in \mathbb{R}^{1\times T \times F}$ (with $\alpha=0.3$) and the complex Cartesian projection of the phase spectrum $\mathbf{P} = \cos \mathbf{Y}_p + i \sin \mathbf{Y}_p \in \mathbb{C}^{1\times T \times F}$ as the magnitude and phase stream inputs, respectively, where $\mathbf{Y}$ denotes the complex spectrum of degraded speech, $T$ and $F$ denote the time and frequency dimensions of the STFT. This initial layer projects the raw inputs into high-dimensional feature spaces $\mathbb{R}^{C_{\text{mag}}\times T \times F}$ and $\mathbb{C}^{C_{\text{pha}}\times T \times F}$, expanding the channel dimensions to $C_{\text{mag}}$ and $C_{\text{pha}}$ for the magnitude and phase streams, respectively. Following this expansion, the features enter a \change{Magnitude-Phase} Dilated DenseNet, containing four MPICMs, which processes phase and magnitude features concurrently. Since dense connectivity inherently involves concatenating all preceding feature maps\change{\cite{pandey2020densely}}, we explicitly re-organize the order of accumulated channels from the two streams, ensuring that the magnitude and phase channels are strictly aligned for the subsequent MPICMs, as illustrated in Fig. \ref{fig:wang1} (b). A final MPICM performs down-sampling, reducing the frequency dimension of both streams to $F' = F/2$ to reduce computational cost of the subsequent bottleneck.
	
	\subsubsection{Bottleneck}\label{bottleneck_overview}
	To capture long-range contextual dependencies, the latent features of each stream are processed by a bottleneck comprising $N=4$ cascaded Dual-Path blocks, \change{where the number of cascaded blocks is ported from \cite{abdulatif2024cmgan, lu2023mp}}. Each dual-path block consists of two serially connected sub-modules: a Time-HADF and a Frequency-HADF. While sharing an identical HADF architecture (detailed in Sec. III-D), these sub-modules operate on \change{time and frequency} dimensions \change{alternately} to capture axis-specific dependencies. Specifically, denoting the input representation with the shape of $B\times {C_{\text{\{mag,pha\}}} \times T \times F'}$ where we added the batch dimension $B$ for illustration. This representation is reshaped to $(B F') \times T \times C_{\{\text{mag,pha}\}}$ and processed by the Time-HADF to model temporal dependencies. Subsequently, the features are permuted to $(B T)\times F'\times C_{\{\text{mag,pha}\}}$, allowing the Frequency-HADF to capture global spectral dependencies. This alternating factorization allows the model to learn full-context spectro-temporal interactions efficiently.  \change{Following the flow diagram illustrated in Fig. \ref{fig:wang1}c, the outputs of time and frequency HADF are reshaped to the original dimension and fused with the block's initial input via a global residual connection, preserving the original signal information while integrating the learned transformations.} 
	
	\subsubsection{Decoder} The decoder reconstructs the signal from the bottleneck output. It employs a \change{Magnitude-Phase} Dilated DenseNet with the same structure as in the encoder, \change{but} operates on the down-sampled resolution, followed by an up-sampling MPICM that restores the original frequency dimension $F$. The final layers employ real and complex convolutions to get the estimated compressed magnitude and phase spectra from their respective streams. Specifically, the predicted magnitude is passed through a ReLU activation to enforce non-negativity, while the final phase angle is recovered by applying the arctangent function to the decoded complex phase features.

	\subsection{Magnitude-Phase Interactive Convolution Module}
	\label{subsection:mpicm}
	The MPICM is designed to process magnitude and phase features in parallel while enforcing geometric constraints. It accepts magnitude $\mathbf{M}_{\text{in}}$ and phase $\mathbf{P}_{\text{in}}$ inputs with channel dimension $C_{\text{mag}}, C_{\text{pha}}$ respectively and spatial dimensions $(T, K)$, where $K$ adapts to the operating frequency resolution. The workflow is illustrated in Fig. \ref{fig:wang3}, which comprises two stages:
	\begin{figure}[t]
		\centering
		\includegraphics[width=\columnwidth]{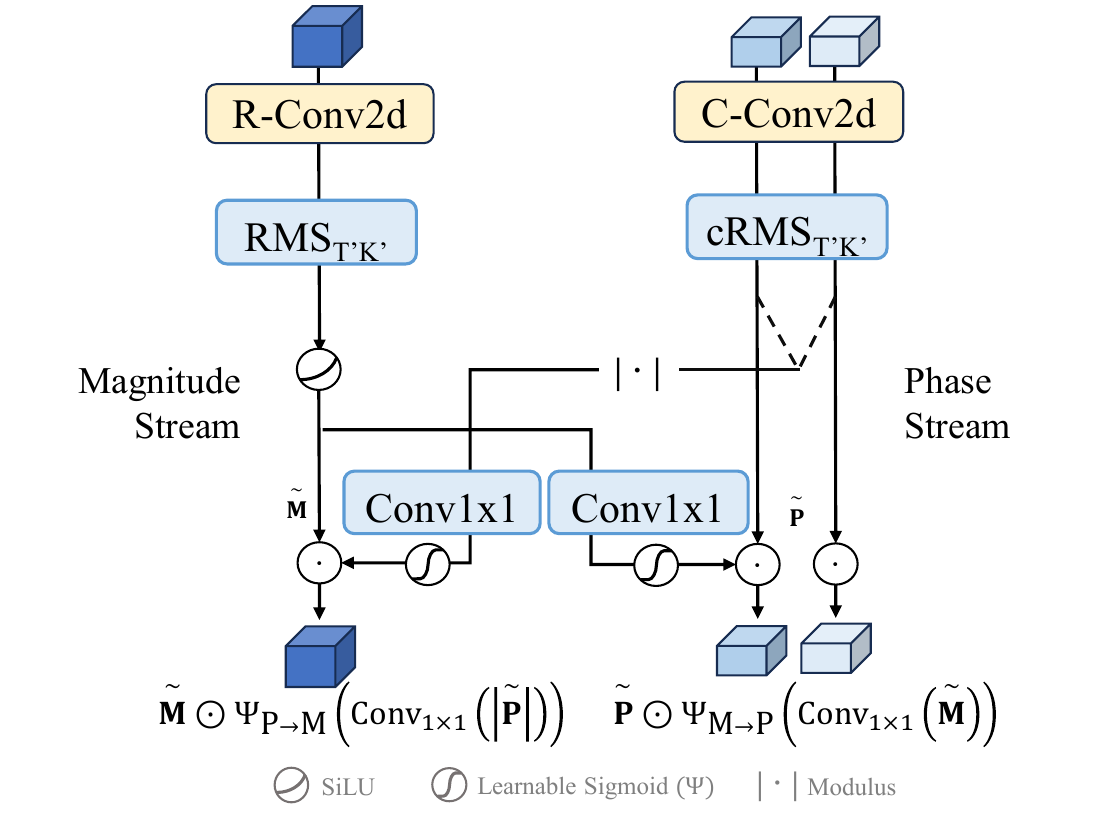}
		\caption{Detailed structure of the MPICM block, including the magnitude and phase dual-streams and their interaction via the gating mechanism.}
		\label{fig:wang3}
	\end{figure}
	\subsubsection{Parallel Feature Extraction}We extract intra-stream features using distinct architectural paradigms to respect the intrinsic structure of each modality.
	
	\textbf{Phase Stream:} To ensure the network learns intrinsic geometric structures, we impose strict rotation equivariance by designing a bias-free and activation-free pipeline. The reason for this design choice is that a fixed bias vector defines a preferred direction in the complex plane, breaking the symmetry required for rotation equivariance, and common complex-valued activations (e.g., ModReLU) often introduce training instability due to undefined gradient regions\change{\cite{lee2022complex}}. We rely on the subsequent gating stage for non-linearity. The intermediate feature is computed as:
	\begin{equation}
		\tilde{\mathbf{P}} = \text{cRMS}_{\text{T'K'}}\left( \text{ComplexConv}(\mathbf{P}_{\text{in}}) \right).
	\end{equation}
	Here, $\text{ComplexConv}$ denotes a bias-free complex-valued convolution, mapping the input to the shape of $ \mathbb{C}^{C_{\text{out}}^{\text{pha}} \times T' \times K'}$.  In our usage of MPICM, temporal resolution is strictly preserved (i.e., $T'=T$), while the output frequency dimension $K'$ varies according to the module's specific function (maintaining $K'=K$ for standard layers, or adjusting appropriately for up-sampling or down-sampling operations).  $\text{cRMS}_{\text{T'K'}}$ represents Complex RMS Normalization \cite{lee2022complex} computed over the joint spatial dimensions $(T', K')$, followed by a frequency-channel-aware learnable scaling vector $\mathbf{\gamma} \in \mathbb{R}^{C_{\text{pha}}^{\text{out}} \times 1 \times K'}$ to capture band-specific importance.
	
	\textbf{Magnitude Stream:} Magnitude spectrum, which directly represents the signal's energy distribution, is naturally suited for conventional neural network pipelines. \change{As a result, for the magnitude stream, we design a convolution-norm-activation block using RMS Normalization\cite{zhang2019root}  followed by SiLU activation}:
	\begin{equation}
		\tilde{\mathbf{M}} = \text{SiLU}\left( \text{RMS}_{\text{T'K'}}\left( \text{Conv2d}(\mathbf{M}_{\text{in}}) \right) \right),
	\end{equation}
	The kernel size, stride, padding and dilation settings of the convolution are identical to that being used in the phase stream, resulting in the feature in $\mathbb{R}^{C_{\text{mag}}^{\text{out}} \times T' \times K'}$. \change{The} $\text{RMS}_{\text{T'K'}}$ here is the real-valued RMSNorm computed over spatial dimensions, and includes standard learnable affine parameters (scale $\gamma$ and bias $\beta$ both in $\mathbb{R}^{C_{\text{mag}}^{\text{out}} \times 1 \times 1}$) applied channel-wise.
	\subsubsection{Interactive Gating}
	To enable cross-modal information flow without violating the rotation equivariance of the phase stream, we propose an interactive gating mechanism that operates element-wise on the extracted features $\tilde{\mathbf{M}}$ and $\tilde{\mathbf{P}}$. The phase-to-magnitude gate is derived from the rotation-invariant modulus $|\tilde{\mathbf{P}}|$, while the magnitude-to-phase gate is derived directly from $\tilde{\mathbf{M}}$. The interaction is defined as:
	\begin{equation}
		\begin{aligned}
			\mathbf{M}_{\text{out}} &= \tilde{\mathbf{M}} \odot \Psi_{\text{P} \to \text{M}}\left( \text{Conv}_{1 \times 1}\left( |\tilde{\mathbf{P}}| \right) \right),\\
			\mathbf{P}_{\text{out}} &= \tilde{\mathbf{P}} \odot \Psi_{\text{M} \to \text{P}}\left( \text{Conv}_{1 \times 1}\left( \tilde{\mathbf{M}} \right) \right).
		\end{aligned}
	\end{equation}
	where $\text{Conv}_{1 \times 1}$ projects the channel dimensions to match the target stream  (e.g., $C_{\text{pha}}^{\text{out}} \to C_{\text{mag}}^{\text{out}}$). $\Psi(\cdot)$ denotes a frequency-channel adaptive gating function:
	\begin{equation}\Psi(\mathbf{X}) = \alpha \cdot \text{sigmoid}(\mathbf{A} \odot \mathbf{X}),
	\end{equation}where $\mathbf{A} \in \mathbb{R}^{C_{\text{\{mag,pha\}}}^{\text{out}}\times1\times  K'}$ is a learnable weight tensor allowing channel-frequency-specific calibration, and $\alpha=3$ is a fixed scaling factor to extend the dynamic range.
	
	 \change{This design makes the global rotation in $\mathbf{P}_{\text{in}}$ rotate $\tilde{\mathbf{P}}$ but leave the magnitude gate unchanged. Consequently, the only effect on the entire block is a global rotation of the output phase feature by the same angle.} It should be noticed that the initial MPICM block which expands the magnitude and phase features omits this gating interaction, as the raw input features do not yet possess the abstract-level information required for cross-stream interaction.

	\subsection{Hybrid-Attention Dual \change{Feed-Forward Network} (HADF)}
	\label{subsection:hadf}
	The HADF block modulates feature importance via joint magnitude-phase contexts and subsequently activates them using stream-specific non-linearities.
	\subsubsection{Signal Flow and Structure}
	\begin{figure*}[t]
		\centering
		\includegraphics[width=\textwidth]{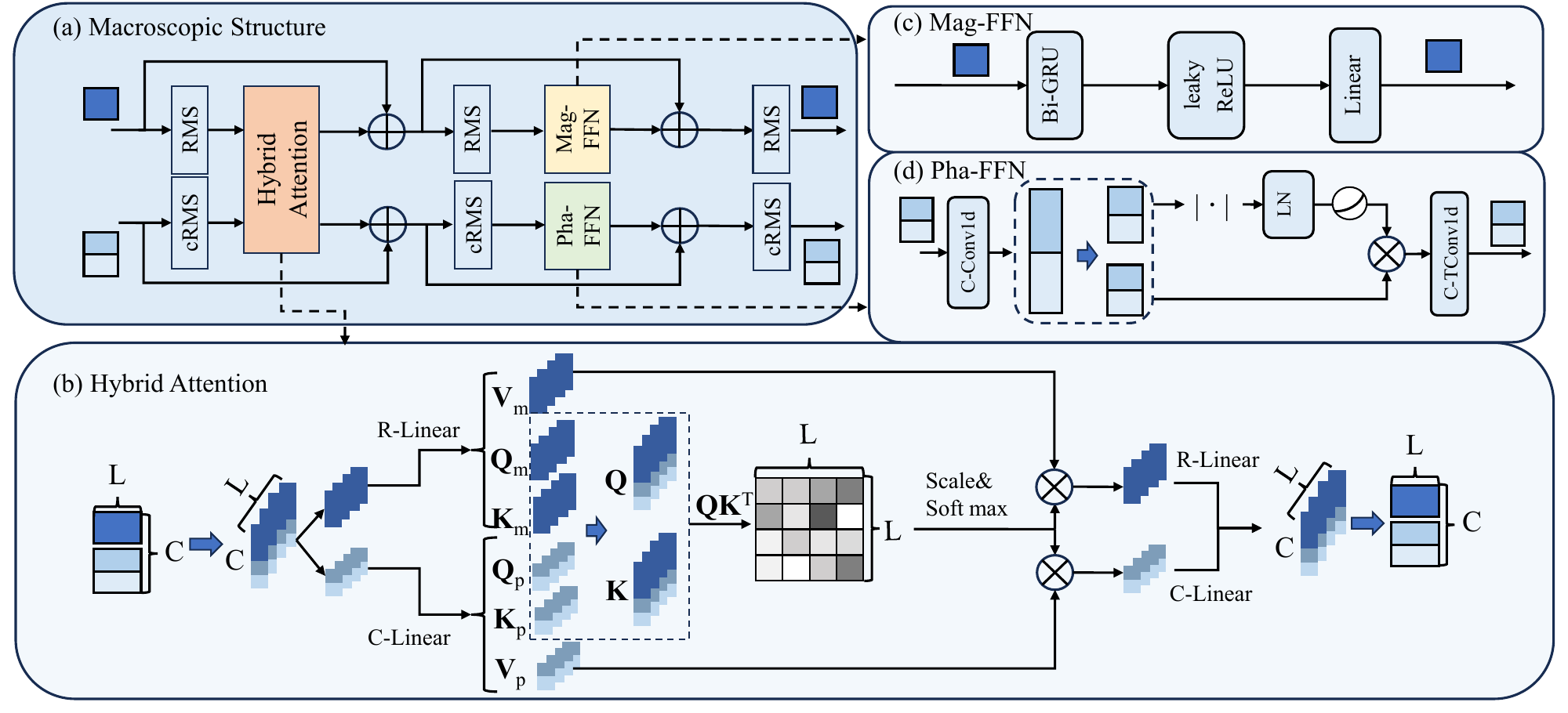}
		\caption{Detailed architecture of the Hybrid-Attention Dual \change{Feed-Forward Network} (HADF) module. (\change{a}) The macroscopic block structure. (\change{b}) The Hybrid Attention mechanism, illustrating the projection of complex queries/keys into a unified attention map. \change{(c) The feed-forward network structure for the magnitude stream (Mag-FFN). (d) The feed-forward network structure for the phase stream (Pha-FFN).}}
		\label{fig:wang9}
	\end{figure*}
	Let $\mathbf{Z}_{\text{mag}} \in \mathbb{R}^{B' \times L \times C_{\text{mag}}}$ and $\mathbf{Z}_{\text{pha}} \in \mathbb{C}^{B' \times L \times C_{\text{pha}}}$ denote the intermediate bottleneck features, where the effective batch size $B'$ absorbs the orthogonal spatial dimension. \change{As in Fig. \ref{fig:wang9}a,} the block processes these features through two residual sub-layers: a Hybrid Attention layer and a Dual \change{Feed-Forward Network (Dual-FFN)} layer. Both streams employ Pre-Normalization, utilizing standard RMS Norm for magnitude and Complex RMS Norm (cRMS) for phase. The macroscopic signal flow is defined as:
	\begin{equation}
		\begin{aligned}
			\mathbf{H}_\text{mag}, \mathbf{H}_{\text{pha}} &= \text{HybridAttention}(\text{RMS}({\mathbf{Z}}_\text{mag}), \text{cRMS}({\mathbf{Z}}_{\text{pha}})), \\
			\mathbf{Z}_\text{mag}, \mathbf{Z}_{\text{pha}} &\leftarrow \mathbf{Z}_\text{mag} + \mathbf{H}_\text{mag}, \mathbf{Z}_{\text{pha}} + \mathbf{H}_{\text{pha}}\\
			\mathbf{F}_\text{mag}, \mathbf{F}_\text{pha} &= \text{Mag-FFN}(\text{RMS}({\mathbf{Z}}_\text{mag})), \text{Pha-FFN}(\text{cRMS}({\mathbf{Z}}_{\text{pha}})),  \\
			\mathbf{Z}_\text{mag}, 	\mathbf{Z}_{\text{pha}} &\leftarrow  \text{RMS}(\mathbf{Z}_\text{mag} + \mathbf{F}_\text{mag}), \text{cRMS}(\mathbf{Z}_{\text{pha}} + \mathbf{F}_{\text{pha}}).
		\end{aligned}
		\label{eq:hadf_flow}
	\end{equation}
	where $\mathbf{H}$ and $\mathbf{F}$ denote the outputs of hybrid attention, and FFN layers respectively, both maintaining the same tensor dimensions as the input $\mathbf{Z}$ in each stream. 	
	\subsubsection{Hybrid Attention Mechanism}
	Effective fusion of magnitude and phase presents a unique challenge: the two streams describe the same signal yet resides on incompatible manifolds. To address this, we propose a Hybrid Attention strategy designed to fuse information in the attention score domain rather than the feature domain\change{, illustrated in Fig. \ref{fig:wang9}b}.
	
	\textbf{Feature Projection}: \change{To satisfy rotation equivariance in the phase stream}, we generate the Query ($\mathbf{Q}$), Key ($\mathbf{K}$), and Value ($\mathbf{V}$) projections for each stream independently\change{, using 4 attention heads following the configuration in \cite{lu2023mp}}. For a given head $h$, these are computed via \textit{bias-free} linear transformations\change{:}
	\begin{equation}
		\begin{aligned}
			\mathbf{Q}_{\text{mag}}^{(h)}, \mathbf{K}_{\text{mag}}^{(h)}, \mathbf{V}_{\text{mag}}^{(h)} &= \text{Linear}_{{Q,K,V}}^{(h)}(\mathbf{Z}_{\text{mag}}),\\ \mathbf{Q}_{\text{pha}}^{(h)}, \mathbf{K}_{\text{pha}}^{(h)}, \mathbf{V}_{\text{pha}}^{(h)} &= \text{ComplexLinear}_{{Q,K,V}}^{(h)}(\mathbf{Z}_{\text{pha}}).
		\end{aligned}
	\end{equation}
	where the projected feature sets lie in $\mathbb{R}^{B'\times L\times C_{\text{mag\_head}}}$ and $\mathbb{C}^{B'\times L\times C_{\text{pha\_head}}}$, respectively, with $C_{\text{mag\_head}}, C_{\text{pha\_head}}$ representing the magnitude and phase channel depth allocated to each attention head, respectively.
	
	\textbf{Unified Scoring:} To derive a shared attention map reflecting the joint confidence of both \change{streams}, we fuse the queries and keys \change{by concatenating them}. For each head $h$, the complex phase vectors are decomposed and concatenated with the magnitude vectors:
	\begin{equation}
		\mathbf{Q}^{(h)} = \text{Concat}\left[ \mathbf{Q}_{\text{mag}}^{(h)}, \text{Re}(\mathbf{Q}_{\text{pha}}^{(h)}), \text{Im}(\mathbf{Q}_{\text{pha}}^{(h)}) \right],
		\label{eq:q_concat}
	\end{equation}
	and similarly for $\mathbf{K}^{(h)}$. The shared attention score $\mathbf{S}^{(h)}$ is computed via the standard scaled dot-product:
	\begin{equation}
		\mathbf{S}^{(h)} = \text{Softmax}\left( \frac{\mathbf{Q}^{(h)}(\mathbf{K}^{(h)})^\top}{\sqrt{d_k}} \right),
	\end{equation}
	where $\mathbf{S}^{(h)}\in \mathbb{R}^{B\times L\times L}$, $d_k = C_{\text{mag\_head}} + 2C_{\text{pha\_head}}$ is the dimension of the concatenated vector per head. 
	Notably, this operation mathematically preserves rotation invariance. The contribution of the phase stream to the dot product corresponds to the real part of the Hermitian inner product \cite{eilers2023building}:
	\begin{equation}
		\text{Re}(\mathbf{Q}_{\text{pha}}) \cdot \text{Re}(\mathbf{K}_{\text{pha}})^T + \text{Im}(\mathbf{Q}_{\text{pha}}) \cdot \text{Im}(\mathbf{K}_{\text{pha}})^T = \text{Re}(\mathbf{Q}_{\text{pha}} \mathbf{K}_{\text{pha}}^\mathcal{H}).
	\end{equation}
	Since $\text{Re}( (\mathbf{q}e^{j\theta}) (\mathbf{k}e^{j\theta})^\mathcal{H} ) = \text{Re}( \mathbf{q}\mathbf{k}^\mathcal{H} )$, global phase rotations cancel out, ensuring the attention scores remain invariant.
	
	\textbf{Output Projection:} The \change{rotation-}invariant scores $\mathbf{S}^{(h)}$ modulate the independent Value vectors $\mathbf{V}_{\text{mag}}^{(h)}$ and $\mathbf{V}_{\text{pha}}^{(h)}$ separately. The head outputs are concatenated and projected back to the original stream dimensions via stream-specific output projections $\text{Linear}_O$ and $\text{ComplexLinear}_O$ (where the latter is bias-free), generating the final residuals:
	\begin{equation}
		\begin{aligned}
			\mathbf{H}_\text{mag} &= \text{Linear}_O \left( \text{Concat}_{h} (\mathbf{S}^{(h)} \mathbf{V}_{\text{mag}}^{(h)}) \right), \\
			\mathbf{H}_\text{pha} &= \text{ComplexLinear}_O \left( \text{Concat}_{h} (\mathbf{S}^{(h)} \mathbf{V}_{\text{pha}}^{(h)}) \right).
		\end{aligned}
	\end{equation}
	\subsubsection{Dual-FFN}
	We employ a Convolution-based FFN for the phase stream to prioritize local dependency modeling, and a GRU-based FFN for the magnitude stream to capture global sequential relationships. This design is premised on the insight that phase information governs fine-grained temporal alignment and local waveform consistency, whereas magnitude information predominates in defining the \change{holistic} spectral envelope and long-range semantic structure.

	\textbf{Magnitude Stream:} To capture \change{long-range} sequential dependencies,  we adopt a GRU-based architecture as in \cite{wang2021tstnn, lu2025explicit}. \change{Depicted in Fig. \ref{fig:wang9}c}, this submodule consists of a bidirectional GRU layer that expands the original feature dimension $C_{\text{mag}}$ to the hidden dimension $C_{\text{mag\_hidden}}$ to capture full-sequence context, followed by a Leaky ReLU activation for non-linearity, and a final linear projection to restore the original channel dimension $C_{\text{mag}}$.
	
	\textbf{Phase Stream}:
	To capture local geometric structures while enforcing rotation equivariance, we design a complex Gated Linear Unit network\change{, as illustrated in Fig. \ref{fig:wang9}d}. The features are first expanded via a complex convolution and subsequently split into two equal halves, $\mathbf{Z}_1$ and $\mathbf{Z}_2$. Both tensors reside in the complex space $\mathbb{C}^{B' \times L \times C_{\text{pha\_hidden}}}$, where $C_{\text{pha\_hidden}}$ denotes the expanded channel width. The gating factor is derived solely from the magnitude of $\mathbf{Z}_2$, which is then normalized and activated via SiLU to produce a real-valued mask, and scales the complex features of $\mathbf{Z}_1$. Finally, a complex projection layer restores the features to original dimension $C_{\text{pha}}$. We denote the sequential processing of the phase stream input $\mathbf{Z}$ as (\ref{eq:ffn phase stream}):
	\begin{equation}
		\begin{aligned}
			[\mathbf{Z}_1, \mathbf{Z}_2] &= \text{Split}(\text{ComplexConv1D}({\mathbf{Z}}_{\text{normed\_pha}})),\\
			\mathbf{F}_{\text{pha}} &= \text{ComplexDeConv}\left( \mathbf{Z}_1 \odot \text{SiLU}(\text{LayerNorm}(|\mathbf{Z}_2|)) \right).
		\end{aligned}
		\label{eq:ffn phase stream}
	\end{equation}
	where $\odot$ denotes element-wise multiplication. Since the gate is derived from rotation-invariant moduli, the rotation of the input $\mathbf{Z}$ ($e^{j\theta}$) propagates linearly through $\mathbf{Z_1}$ and is preserved at the output, satisfying the equivariance condition.
	
	\section{Experimental Setup}
	To comprehensively validate the proposed method, we adopt a hierarchical evaluation strategy. We begin by isolating the model's intrinsic phase modeling capabilities through a dedicated Phase Retrieval task, where the model is provided with the clean compressed magnitude spectrum while the input phase is strictly initialized to zero. After that, we benchmark denoising performance on the standard publicly available VoiceBank+DEMAND \cite{botinhao2016investigating} and Deep Noise Suppression (DNS) Challenge 2020 \cite{reddy20interspeech} datasets. We also perform zero-shot cross-dataset test to verify the generalization ability.  Furthermore, to investigate the relative importance of our equivariant phase modeling across various types of degradations, we construct a custom training set derived from DNS 2021 and evaluate performance on a simulated out-of-domain WSJ0+WHAMR! test set and perform ablation studies on it. This custom dataset covers extensive degradations across three representative distortions: denoising, dereverberation, and bandwidth extension, as well as their combinations.
	\subsection{Datasets}
	\subsubsection{VoiceBank+DEMAND}
	The VoiceBank+DEMAND (VBD) corpus is a typical benchmark for speech denoising, constructed by mixing clean speech from the Voice Bank corpus \cite{veaux2013voice} with diverse acoustic environments from the DEMAND database \cite{thiemann2013diverse} and artificial noise sources. All audio recordings were downsampled to 16 kHz in our experiment.
	The training set comprises 11,572 utterances from 28 speakers. We also perform the phase retrieval task on the clean speech partition (VoiceBank corpus \cite{veaux2013voice}), adhering to the standard speaker-disjoint split.
	
	\subsubsection{DNS Challenge 2020} The DNS Challenge-2020 corpus \cite{reddy20interspeech} provides over 500 hours of high-quality clean speech from 2,150 distinct speakers and approximately 180 hours of diverse noise clips. Following the official script, we generated a total of 3,000 hours of noisy-clean training pairs, with SNR levels of noisy speech randomly sampled from a uniform distribution between -5 dB and 15 dB, without reverberation. For evaluation, we utilize the non-reverberant category for official non-blind test set, which consists of 150 synthetic recordings.
	
	\subsubsection{DNS-Challenge 2021 + WSJ0-WHAMR!}
	We adopt the DNS Challenge 2021 corpus \cite{reddy2021icassp} as the training set for our comprehensive speech restoration experiments, which provides a significantly richer set of room impulse responses.
	The clean speech is exclusively sourced from the read\_speech subset of DNS 2021. To strictly evaluate out-of-domain generalization, the test set is derived from the si\_et\_05 evaluation subset of WSJ0 \cite{garofolo1993csri}, comprising 651 recordings from eight distinct speakers. Our simulation protocol covers three representative distortion categories:
	
	\textbf{Denoising (DN):} Training samples are generated by mixing clean speech with noise clips randomly sampled from the noise in DNS-2021 corpus at SNRs uniformly distributed between -5 dB and 15 dB. To ensure out-of-domain evaluation, we employ noise recordings exclusively from the WHAMR! \cite{maciejewski2020whamr} test partition. This subset comprises 3,000 distinct clips captured in diverse urban environments (e.g., cafes, restaurants, and bars). We randomly mix them with clean speech at fixed SNR levels of \change{-5,} 0, 5, 10, and 15 dB.

	\textbf{Dereverberation (DR):} 
	For the training set, we simulate reverberant conditions by convolving clean speech with Room Impulse Responses (RIRs) randomly sampled from the DNS-2021 dataset. For the test set, reverberation is generated using the official WHAMR! simulation script based on the image-source method \cite{allen1979image}. 
	
	\textbf{Bandwidth Extension (BWE):}
	To simulate bandwidth limitations, we employ a resampling-based strategy. The input signals are downsampled to a target effective sampling rate and subsequently upsampled back to the original resolution (16 kHz). For standalone BWE tasks, we generate samples with effective cutoff frequencies of 2 kHz and 4 kHz.
	
	\textbf{Composite Distortions:}
	To simulate complex acoustic environments, we combine the standalone tasks into two composite scenarios: DN+DR, and DN+DR+BWE. The noise and reverberation settings remain consistent with their standalone definitions, while the bandwidth extension component is restricted to a 4 kHz cutoff in mixed scenarios to preserve essential spectral cues \cite{liu2025universal}.
	
	The final dataset is balanced uniformly across these categories. The training partition comprises 300 hours of audio. The test set contains 1,250 utterances (250 per category) generated from WSJ0 clean speech. For all noise-inclusive test conditions, samples are evenly distributed across five SNR levels as previously described (-5 to 15 dB in 5 dB intervals), with 50 files allocated to each level.
	
	\subsection{Model and Training Configurations}
	\label{subsection:model and training configurations}
	For STFT configuration, a window size of 25 ms is adopted, with 25\% shift between adjacent frames, resulting in $F=201$ frequency bins. We design two model variants, which is Small for the \change{Phase Retrieval (PR)} task and Standard for the others, with specific channel configurations detailed in Table \ref{tab:configuration}.  Given that complex-valued phase channels require approximately double the computational resources of real-valued magnitude channels, we maintain the phase channel number at roughly half of the magnitude one. We will further investigate the impact of this allocation strategy via ablation studies in Section IV-D.

	\begin{table}[htbp]
	\centering
	\caption{Channel configurations for the proposed model variants.}
	\small
	\resizebox{\columnwidth}{!}{%
		\begin{tabular}{lcccccc}
			\toprule
			Model & $C_\text{mag}$ & $C_\text{pha}$ & $C_\text{mag\_head}$ & $C_\text{pha\_head}$ & $C_\text{mag\_hidden}$ & $C_\text{pha\_hidden}$ \\
			\midrule
			Small  & 32 & 16 & 8  & 6  & 64 & 64 \\
			Standard & 48 & 16 & 12 & 6  & 96 & 64 \\
			\bottomrule
		\end{tabular}%
	}
	\label{tab:configuration}
\end{table}
	
	We adopt the typical  loss \change{terms} from MP-SENet \cite{lu2023mp} for the pure DN task. This objective $\mathcal{L}_{\text{DN}}$ aggregates magnitude, complex, time-domain, STFT consistency, and PESQ metric discriminator losses with a phase term $\mathcal{L}_{\text{pha}}$ which considers group delay, instantaneous phase and angular frequency  \cite{ai2023neural}.
	\change{\begin{equation} \label{eq:dn_loss}
		\mathcal{L}_{\text{DN}} = \lambda_{1}\mathcal{L}_{\text{mag}} + \lambda_{2}\mathcal{L}_{\text{pha}}  + \lambda_{3}\mathcal{L}_{\text{com}} + \lambda_{4}\mathcal{L}_{\text{Metric}} + \lambda_{5}\mathcal{L}_{\text{con}} + \lambda_{6}\mathcal{L}_{\text{time}}
	\end{equation}
For fair comparison, all predictive methods evaluated in this study were trained with the identical loss configuration and weights \cite{lu2025explicit}, with $\lambda_1=0.9$, $\lambda_2=0.3$, $\lambda_3=0.2$, $\lambda_4=0.05$, $\lambda_5=0.1$, $\lambda_6=0.2$.}
	To address the enhancement failure of conventional methods in BWE-involved universal SE tasks \cite{liu2025universal}, we augment the objective with a Multi-Period Discriminator (MPD) \cite{kong2020hifi, lu2025towards} for all evaluated models. Accordingly, the loss function for the USE task is defined as: 
	\change{\begin{equation} \label{eq:use_loss} \mathcal{L}_{\text{USE}} = \mathcal{L}_{\text{DN}} + \lambda_{\text{mpd}} \mathcal{L}_{\text{MPD}} \end{equation}with $\lambda_{\text{mpd}}=  0.05$.} For the Phase Retrieval task, we switch to a loss term better suited for synthesizing phase from scratch. Following the loss  setting in BAPEN \cite{dai2025bapen}, this specific loss $\mathcal{L}_{\text{PR}}$ emphasizes geometric consistency via an omni-directional phase distortion term and an MPD adversarial term:
	\change{
	\begin{equation}
		\mathcal{L}_{\text{PR}} = \lambda_{\text{o}} \mathcal{L}_{\text{omni}} + \lambda_{\text{m}}\cdot \mathcal{L}_{\text{MPD}}
	\end{equation}
	}\change{with $\lambda_{\text{o}}=2\cdot 10^4$, $\lambda_{\text{m}} = 1$ which are consistent with the settings in \cite{dai2025bapen}.}
	
	While loss configurations varied by task, all models were trained using a batch size of $B=4$ and the AdamW optimizer ($\beta_1 = 0.8, \beta_2 = 0.99$, weight decay $0.01$). We set the initial learning rate to $5 \times 10^{-4}$, employing an exponential decay of 0.99 per epoch for DNS-2020 (1M steps) and VoiceBank+DEMAND (500k steps). For the simulated DNS-Challenge 2021 corpus, we utilized a slower decay of 0.999 over 500k steps to accommodate its complex degradation. 

\subsection{Evaluation Metrics}
\change{
We evaluate performance using a comprehensive suite of metrics. To verify phase reconstruction accuracy, we report the Phase Distance (PD) \cite{choi2019phase} and Weighted Omni-directional Phase Distortion (WOPD) \cite{dai2025bapen}, where lower values denote better performance. PD computes the target‑magnitude‑weighted average of the absolute angular difference between the estimated and clean phase spectra, directly measuring phase deviation. WOPD calculates magnitude‑weighted omni‑directionally consistent phase distortion across nine wrap‑free directions, capturing relative phase differences. Together, these two metrics provide a rigorous evaluation of the proposed method's phase estimation accuracy.}

\change{General restoration quality is assessed via standard reference-based metrics: Perceptual Evaluation of Speech Quality (PESQ) \cite{rix2001perceptual}, Short-Time Objective Intelligibility (STOI) \cite{taal2010short}, Composite Objective Measure for Overall Speech Quality (COVL) \cite{hu2008evaluation}, and Scale-Invariant Signal-to-Distortion Ratio (SI-SDR) \cite{le2019sdr}. PESQ and STOI evaluate perceptual speech quality and speech intelligibility, respectively, while COVL balances various aspects of degradation by combining multiple distortion measures into a single MOS-like score. SI-SDR measures waveform fidelity by comparing the estimated time-domain signal to the clean reference after scaling to remove amplitude ambiguity, which is highly sensitive to phase misalignment. For all general restoration quality metrics, higher values indicate better speech quality.}

\change{To supplement these reference‑based metrics, we also employ two reference‑free estimators, DNSMOS \cite{reddy2021dnsmos} and UTMOS \cite{saeki2022utmos}, both of which are trained to correlate with human perceptual rating, with higher scores indicate better human-perceived quality. DNSMOS is a non-intrusive perceptual metric trained on extensive human ratings, enabling it to evaluate noise suppression algorithms. Additionally, UTMOS provides a reference-free evaluation of overall speech naturalness by utilizing deep self-supervised learning representations to predict human mean-opinion-scores.}
	
	\section{Results and Analysis}
	
	\subsection{Phase Retrieval}
	\label{subsection:phase retrieval}
	To isolate and validate the model's intrinsic phase modeling capabilities, we first evaluate performance on the phase retrieval task.  Given that only the phase needs to be recovered, we employ the Small configuration of our proposed model. The magnitude stream is not discarded here, since the whole phase stream relies on the magnitude one to get information with the input phase all zeros. The difference is that we don't decode the magnitude from the final MPICM in Fig. \ref{fig:wang1}. We benchmark against a diverse set of baselines, including the classical Griffin-Lim algorithm \cite{griffin1984signal} and generative DiffPhase \cite{peer2023diffphase}. We also include MP-SENet \change{Update} \cite{lu2025explicit} \change{(also denoted as MP-SENet Up.)} and SEMamba \cite{chao2024investigation} as strong predictive baselines, adapting them for this task by retaining only the phase decoder. To ensure fair comparison, all predictive models adhere to the consistent STFT and loss configurations described in Section \ref{subsection:model and training configurations}. An exception is made for DiffPhase due to its generative nature, where its original loss and STFT formulation are maintained \cite{peer2023diffphase}. 
	Quantitative results are presented in Table \ref{tab:phase_retrieval_results}.   As observed, despite utilizing only 0.90M parameters and 22.89 GMACs, our method surpasses all baselines in phase estimation precision (PD and WOPD) by a substantial margin, while achieving a marginal improvement in perceptual quality. This result is expected, given that metrics such as PESQ are relatively insensitive to fine-grained phase alignment. Nevertheless, this task serves as a critical benchmark for isolating intrinsic phase modeling capabilities. We attribute the significant increase in phase accuracy to the rotation-equivariant design, which explicitly constrains the high-dimensional latent space to consider the circular topology of phase. In contrast, conventional predictive approaches expend capacity to implicitly learn this geometric structure, potentially limiting their performance.
	
	\begin{table}[h]
		\centering
		\caption{Comparison of Phase Retrieval performance on the VoiceBank corpus.  $^{*}$ denotes baselines adapted with a single phase decoder, and GRE denotes Global Rotation Equivariance.}
		\begin{threeparttable}
			\resizebox{\columnwidth}{!}{%
				\renewcommand{\arraystretch}{1.3} 
				\begin{tabular}{l|cc|cc|cc}
					\hline
					\rule{0pt}{4ex}\textbf{Model} & \textbf{\makecell{Para. \\ (M)}} & \textbf{\makecell{MACs\\ (G/s)}} & \textbf{PESQ} & \textbf{\makecell{SI-SDR \\ (dB)}}& \textbf{\makecell{WOPD\\ ($\downarrow$)}} & \textbf{\makecell{PD\\ ($\downarrow$)}} \\[2ex] \hline
					Griffin-Lim & - & - & 4.23 & -17.07 & 0.342 & 90.07\\
					
					DiffPhase & 65.6 & 3330 & 4.41 & -11.75 & 0.230& 85.66 \\
					MP-SENet Up.$^{*}$ & 1.99 & 38.80 & 4.60 & 14.64 & 0.058 & 11.38 \\ 
					
					SEMamba$^{*}$ & 1.88 & \change{28.39} & 4.59 & 13.63 & 0.059 & 12.46 \\ 
					Proposed (Small) & \textbf{0.90} & 22.89 & \textbf{4.61} & \textbf{16.03} & \textbf{0.044} & \textbf{8.47} \\ 
					
					\hspace{0.3em}-- break MPICM GRE & \textbf{0.90} & 22.89 & \textbf{4.61} & 14.99 & 0.048 & 9.64 \\ 
					\hspace{0.3em}-- break Attn. GRE & \textbf{0.90} & 22.89 & \textbf{4.61} & 15.67 & 0.048 &  9.16\\
					\hspace{0.3em}-- break FFN. GRE & \textbf{0.90} & 22.89 & {4.60} & 14.97 & 0.050 &  9.67\\ 
					\hline
				\end{tabular}%
			}
		\end{threeparttable}
		\label{tab:phase_retrieval_results}
	\end{table}
	To substantiate the hypothesis that GRE has contribution to the improved phase prediction, we designed specific model variants to examine the impact of deliberately breaking this constraint through two distinct modifications: (1) substituting the modulus in MPICM with the sum of real and imaginary part of the phase feature during interactive gating, thereby breaking the GRE property in the encoder and decoder; (2) inverting the sign of the phase query component (\ref{eq:q_concat}) (i.e., $\text{Re}(\mathbf{Q}_{\text{pha}}) \to -\text{Re}(\mathbf{Q}_{\text{pha}})$), which fundamentally disrupts the GRE in the hybrid attention mechanism; and (3) remove the modulus in (\ref{eq:ffn phase stream}) and apply GLU to real and imaginary components separately, which disrupts the GRE in phase FFN. As in Table \ref{tab:phase_retrieval_results}, all modifications resulted in evident degradation in phase accuracy, confirming that strict adherence to rotation equivariance is instrumental for precise phase modeling.

	\subsection{Speech Denoising on Benchmark Datasets}

	For denoising task, the model has to extract the distribution of clean speech from an  unbounded, open set of noise distributions. Unlike the phase retrieval task, denoising presents a scenario where both magnitude and phase inputs are unreliable, particularly in low-SNR regions.
	\begin{table}[t]
		\centering
		\caption{Denoising performance evaluation on the VoiceBank+DEMAND test set and the zero-shot DNS-2020 Non-Reverberant test set. All models were trained exclusively on the VBD corpus.}
		\label{tab:vbd_result}
		\setlength{\tabcolsep}{3pt}
		\resizebox{\columnwidth}{!}{%
			\begin{tabular}{l cc ccccc cc}
				\toprule
				\textbf{Method} & \textbf{\makecell{Para.\\ (M)}} & \change{\textbf{\makecell{MACs\\(G/sec)}}}& \textbf{PESQ} & \textbf{STOI} & \change{\textbf{\makecell{SI-SDR\\(dB)}}}& \textbf{\makecell{UT-\\MOS}} & \textbf{\makecell{DNS-\\MOS}} & \textbf{\makecell{PD\\ ($\downarrow$)}} & \textbf{\makecell{WOPD \\ ($\downarrow$)}} \\
				\midrule
				
				\multicolumn{10}{c}{{{Test Set: VoiceBank+DEMAND}}} \\
				\midrule
				FRCRN           & 6.90 & \change{37.50}& 3.199 & 0.953 & \change{15.984} & 4.009 & \textbf{3.582} & 14.703 & 0.192 \\
				CMGAN           & 1.83   & \change{58.00} & 3.410 & 0.956 & \change{18.659} & 4.051 & 3.556 & \underline{7.309} & 0.167 \\
				DB-AIAT         & 2.81   & \change{34.00} & 3.264 & 0.956 & \change{19.445} & 4.038 & \underline{3.563} & 7.521 & 0.174 \\
				MP-SENet        & 2.05   & \change{38.90} & 3.496 & \underline{0.960} & \change{\textbf{19.938}}& 4.045 & 3.557 & 7.315 & \underline{0.162} \\
				MP-SENet Up.    & 2.26 & \change{43.14}& \textbf{3.604} & \textbf{0.961} & \change{19.435} & \underline{4.064} & \underline{3.563} & 7.403 & 0.163 \\
				SEMamba         & 2.25   &\change{32.73}& 3.564 & \underline{0.960} & \change{19.080} & 4.041 & 3.549 & 7.443 & 0.164 \\
				\textbf{Proposed} & 1.55 &\change{34.97}& \change{\underline{3.598}} & \change{\underline{0.960}}& \change{\underline{19.925}} & \change{\textbf{4.082}} & \change{3.552} & \change{\textbf{7.212}} & \change{\textbf{0.160}} \\
				
				\midrule
				
				\multicolumn{10}{c}{{{Test Set: DNS-2020 Non-Reverberant}}} \\
				\midrule
				CMGAN           & 1.83   & \change{58.00} & 2.759 & 0.957 & \change{15.721} & 3.675 & \underline{3.941} & 8.490 & 0.190 \\
				MP-SENet        & 2.05   & \change{38.90} & 2.719 & \underline{0.960} & \change{16.126} & \underline{3.697} & 3.930 & 8.423 & 0.185 \\
				MP-SENet Up.    & 2.26 & \change{43.14}& \underline{2.790} & 0.959 & \change{\underline{16.277}}& 3.591 & 3.897 & \underline{8.271} & \underline{0.181} \\
				SEMamba         & 2.25   & \change{32.73} & 2.440 & 0.938 & \change{13.253} & 3.438 & 3.862 & 9.455 & 0.204 \\
				\textbf{Proposed} & 1.55 & \change{34.97} & \change{\textbf{2.920}} & \change{\textbf{0.968}} & \textbf{\change{17.867}} & \change{\textbf{3.889}} & \change{\textbf{4.009}} & \change{\textbf{7.683}} & \change{\textbf{0.170}} \\
				
				\bottomrule
			\end{tabular}%
		}
	\end{table}
	Table \ref{tab:vbd_result} presents the results for models trained on the VoiceBank+DEMAND dataset. We benchmark against a diverse set of competitive baselines: the complex-valued FRCRN \cite{zhao2022frcrn}, the dual-branch methods DB-AIAT \cite{yu2022dbt} and CMGAN \cite{abdulatif2024cmgan}, and the Mamba-based SEMamba \cite{chao2024investigation}. Additionally, we compare against our direct architectural predecessors, MP-SENet \cite{lu2023mp} and MP-SENet Up. \cite{lu2025explicit}, to isolate the specific performance gains attributed to the dual-stream design \change{with} rotation-equivariant phase modeling. Our method achieves competitive performance, \change{attaining the best UTMOS and phase accuracy among all models, while remaining very close to MP-SENet Update on intrusive metrics}. We attribute this \change{slight} discrepancy to the inherent limitations of the VBD corpus; It is very small-scaled, and recording environment acoustical factors presented in the ``clean'' reference targets can cause intrusive metrics to over-penalize valid enhancements. To rigorously validate generalization capability beyond this limited test set, we performed a cross-corpus evaluation by applying the VBD-trained model directly to the DNS non-reverberant test set. In this zero-shot setting, our method demonstrates significant superiority, surpassing all baselines on all metrics with gains exceeding 0.1 in PESQ, alongside an evident reduction in phase-related metrics. This proves that the proposed method is actually equipped with stronger denoising ability in un-seen acoustical scenarios, which underscores the value of our design, showing that 
	the network is compelled to learn \change{more generalizable} magnitude and phase structures rather than memorizing spurious, dataset-specific patterns, thereby showing stronger robustness in unseen acoustic environments.
	\begin{table}[htbp]
		\centering
		\caption{Comparative evaluation of models trained on the large-scale DNS-2020 dataset and tested on the official Non-Reverberant blind test set.}
		\label{tab:dns_result}
		
		\footnotesize 
		
		\setlength{\tabcolsep}{3pt} 
		\resizebox{\columnwidth}{!}{
			\begin{tabular}{lcccccccc}
				\toprule
				\textbf{Method} & \textbf{\makecell{Para.\\ (M)}} & \textbf{PESQ}  & \textbf{\makecell{STOI}} & \textbf{\makecell{SI-SDR}} & \textbf{\makecell{UT-\\MOS}} & \change{\textbf{\makecell{DNS- \\MOS}}} &  \textbf{\makecell{PD\\ ($\downarrow$)}} & \change{\textbf{\makecell{WOPD \\ ($\downarrow$)}}} \\
				\midrule
				Noisy &  -- & 1.582  & 0.915 & 9.230 & 2.365 & \change{3.156} & 9.333 & \change{0.228} \\
				\midrule
				FRCRN &  6.90 & 3.233 & 0.977 & 19.783 & 3.911 & \change{4.039} & 7.432 & \change{0.169} \\
				MFNet &  -- & 3.431 & 0.980 & 20.310  & 4.051 &\change{\textbf{4.098}} & 7.122 & \change{0.164}\\
				MP-SENet Up. &  2.26 & 3.624 & \textbf{0.982} & 21.033 & 4.053 & \change{4.079} & 6.713 & \change{0.151} \\
				SEMamba &  2.85 & 3.593 & 0.981 & 20.732 & 4.024 & \change{4.057} & 6.914 & \change{0.153}\\
				\textbf{Proposed}  &  {1.55} & \textbf{3.645}  & \textbf{0.982} & \textbf{21.060} & \textbf{4.068} & \change{4.078} & \textbf{6.571} & \change{\textbf{0.149}}\\
				\bottomrule
			\end{tabular}
		}
	\end{table}
	Table \ref{tab:dns_result} presents the evaluation results on the large-scale DNS 2020 dataset. Consistent with previous findings, our method surpasses all baselines in phase estimation precision (PD and WOPD)\change{,} simultaneously achieving \change{superior} results in perceptual quality metrics including PESQ and UTMOS\change{, while lagging behind MP-SENet Update by only 0.001 in DNSMOS}.
	This validates that despite the vast scale of the training data, our model (1.55 M parameters) effectively extracts high-fidelity speech distributions, \change{generally} outperforming larger baselines such as MP-SENet Up. ($>$2.2 M parameters). 
	This confirms that appropriate manifold handling with Global Rotation Equivariance (GRE) in phase modeling provides a more effective inductive bias for speech enhancement than simply increasing model capacity.
	
\begin{figure*}[t]
		\centering
		\includegraphics[width=0.8\textwidth]{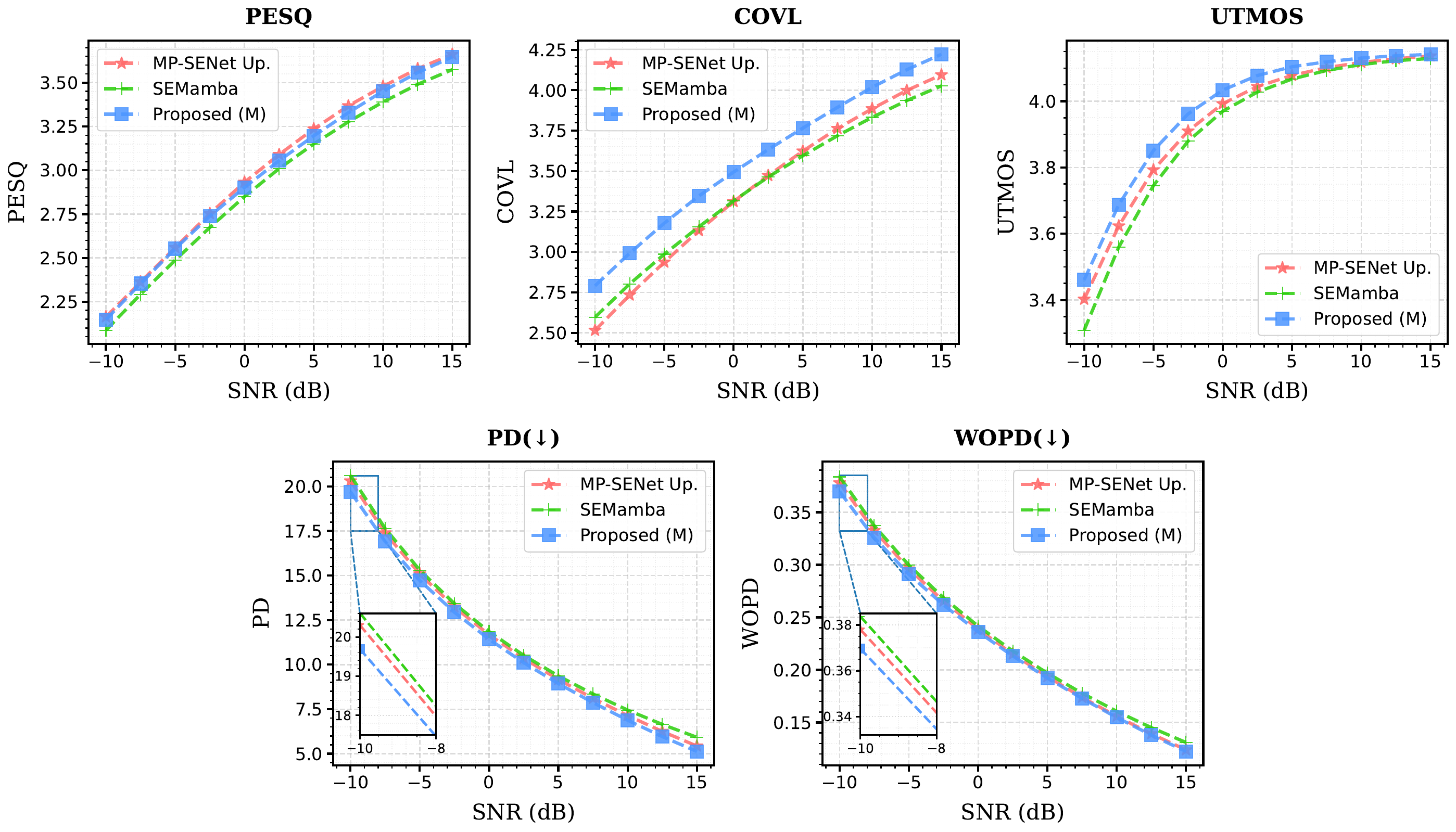}
		\caption{Performance Comparison across varying SNRs. Models were trained on the DNS-2020 corpus and evaluated on re-mixed versions of the VoiceBank+DEMAND test set ranging from -10 dB to 15 dB.}
		\label{fig:metric_per_snr}
	\end{figure*}

	To further investigate the capability of our method under varying acoustic conditions, we also conducted a cross-corpus validation. Specifically, we evaluated the model trained on the DNS dataset using the VBD test set, which was re-mixed at lower SNRs ranging from -10 dB to 15 dB, with the top-ranked SEMamba and MP-SENet Up. for comparison. As illustrated in Fig. \ref{fig:metric_per_snr}, our method demonstrates consistent superiority in low-SNR scenarios, particularly evident in the COVL and UTMOS metrics. 
	The phase-sensitive metrics (PD and WOPD) reveal distinct advantages, despite being relatively close overall. 
	The evident improvements in perceptual quality and the slight improvement in phase accuracy demonstrate the resilience of our proposed method in mismatched and low-SNR environments, highlighting the distinct advantage of our phase modeling over the baselines.
	
	\subsection{Universal Speech Enhancement}
\begin{figure*}[t]
		\centering
		\includegraphics[width=\textwidth]{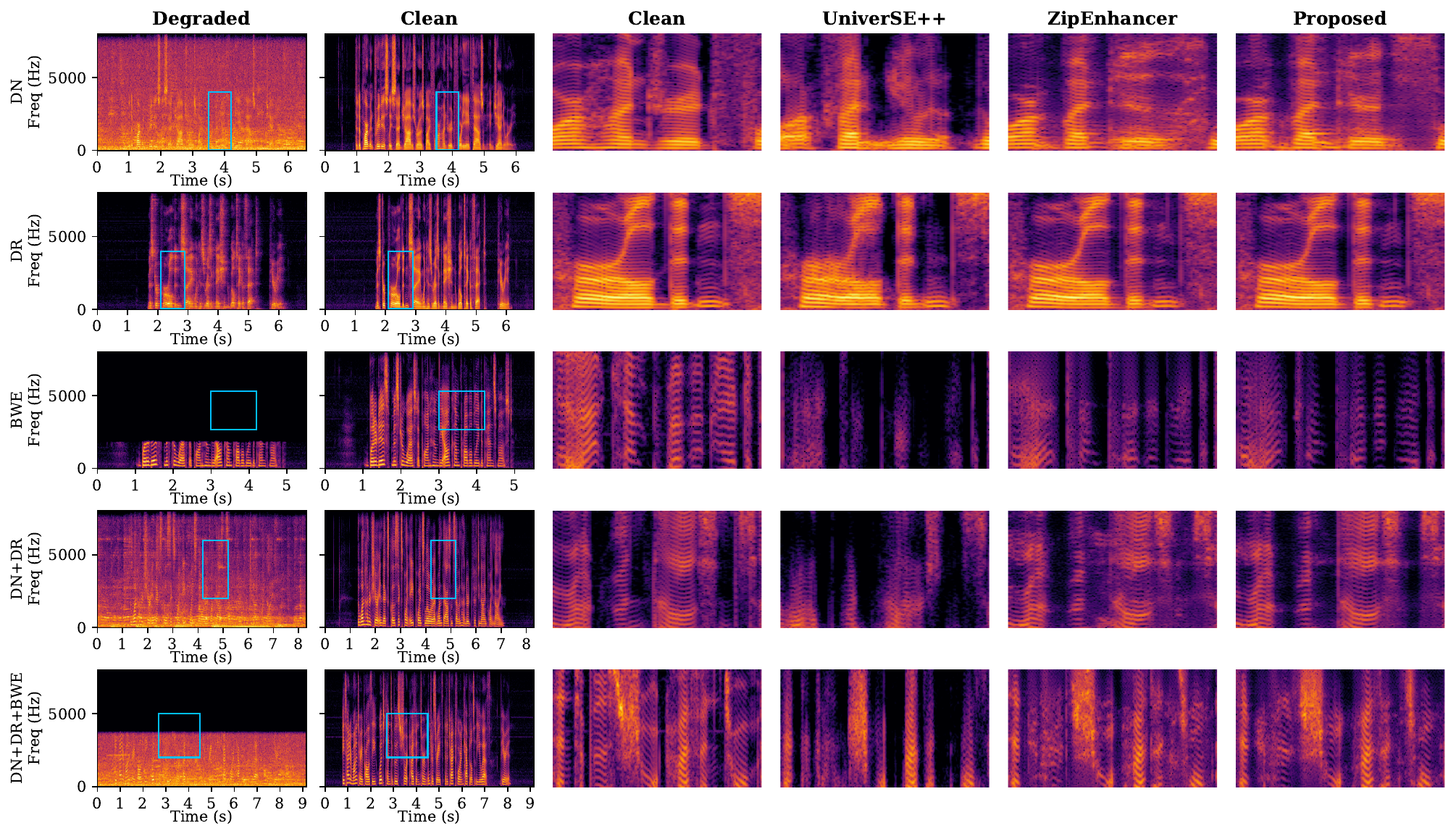}
		\caption{Spectrogram visualization of enhanced speech under diverse distortion scenarios. The audio files are taken from WSJ0+WHAMR! test set.}
		\label{fig:wang6}
	\end{figure*}

	While pure denoising at high SNRs relies predominantly on magnitude recovery, USE tasks involving DR and BWE put more focus on accurate phase reconstruction. 
	In this part, we include the recently proposed ZipEnhancer \cite{wang2025zipenhancer} as another strong baseline, which utilizes multi-resolution resampling to optimize efficiency and performance, as well as the generative diffusion-based UniverseSE++ \cite{scheibler2024}. Since sigmoid-based masking \cite{lu2025explicit} is not fit for the BWE-contained tasks, all predictive baselines are re-trained using a mapping method that employs a ReLU activation to ensure non-negative magnitude output \cite{wang2025zipenhancer}, with consistent training configurations in Section \ref{subsection:model and training configurations}. The params and computational cost comparison is in Table \ref{tab:param_compute_comparison}, quantitative evaluation results are detailed in Table \ref{tab:universal_result}, and spectrogram comparisons are presented in Fig. \ref{fig:wang6}. As in Table \ref{tab:param_compute_comparison}, our method achieves the lowest parameter count, and its computational cost is significantly lower than that of our structural baseline, MP-SENet Up, remaining comparable to distinct architectures like SEMamba and ZipEnhancer. Despite this efficiency, the proposed method demonstrates evident performance gains over these approaches. Readers are encouraged to audit the audio samples and training logs hosted on our project website\footnote{\url{https://wangchengzhong.github.io/GRENet-Supplementary-Materials/}\label{web}}.
	
	\begin{table*}[t]
		\centering
		\caption{Performance comparison for composite SE tasks (Train: DNS-2021, Test: WSJ0+WHAMR!).}
		\label{tab:universal_result}
		\setlength{\tabcolsep}{6pt} 
		
		\begin{tabular}{llcccccc|cc}
			\toprule
			\multirow{2}{*}{\textbf{Task}} & \multirow{2}{*}{\textbf{Method}} & \textbf{WB-PESQ} & \textbf{STOI} & \textbf{SI-SDR} & \textbf{COVL} &  \textbf{UTMOS} & \change{\textbf{DNSMOS}} & \textbf{PD} & \textbf{WOPD} \\
			& & ($\uparrow$) & ($\uparrow$) & ($\uparrow$) & ($\uparrow$) & ($\uparrow$) & \change{($\uparrow$)} & ($\downarrow$) & ($\downarrow$) \\
			\midrule
			
			\multirow{5}{*}{DN} 
			
			& UniverSE++         & 2.007 & 0.935 & 10.795 & 2.531 & 3.708 & \change{3.765} & 23.508 & 0.417 \\
			& MP-SENet Up.   & 2.656 & 0.951 & 14.898 & 3.481 & 3.903 &  \change{\underline{4.004}} & 12.467  & 0.261 \\
			& SEMamba       & 2.658 & 0.950 & 14.311 & 3.427 & \textbf{4.026} & \change{\textbf{4.031}} & 12.559 & 0.256\\
			& ZipEnhancer   & \underline{2.717}& \textbf{0.958} & \textbf{15.251} & \underline{3.507} &  3.896 & \change{4.003} &  \underline{11.862} & \underline{0.250}\\
			& \textbf{Proposed }  & \textbf{2.750} & \underline{0.957} & \underline{15.234} & \textbf{3.545} & \underline{3.945} & \change{4.000} & \textbf{11.857} & \textbf{0.245} \\
			\midrule
			
			\multirow{5}{*}{DR} 
			& UniverSE++          & 2.565 & 0.944 & 6.043 & 2.978 & 3.735 & \change{3.874} & 28.692 & 0.494 \\
			& MP-SENet Up.   & 3.486 & 0.978 & 11.691 & 4.306 & 4.252 & \change{\underline{4.087}} & 12.056 & 0.253 \\
			& SEMamba       & \underline{3.577} & \underline{0.981} & 12.143 & \underline{4.347} & \underline{4.249} & \change{\textbf{4.090}} & 10.548 & \underline{0.221}\\
			& ZipEnhancer   & 3.501 & \underline{0.981} & \underline{12.502} & 4.344 & 4.224 & \change{4.075} & \underline{10.432} & 0.224\\
			& \textbf{Proposed }  & \textbf{3.662} & \textbf{0.987} &  \textbf{13.960} & \textbf{4.472} & \textbf{4.263} & \change{4.042} & \textbf{8.878} & \textbf{0.190}\\
			\midrule
			
			\multirow{5}{*}{BWE} 
			
			& UniverSE++          & 3.257 & 0.954 & 11.243 & 3.413 & 3.943 & \change{3.733} & 24.034 & 0.356 \\
			& MP-SENet Up.   & 3.322 & 0.967 & 11.245 & 3.602 & 4.057 & \change{\textbf{3.954}} & 21.183 & 0.288\\
			& SEMamba       & 3.305 & 0.967 & 10.992 & 3.561 & 4.058 & \change{3.760} & 22.186 & 0.306\\
			& ZipEnhancer   & \underline{3.486} & \textbf{0.971} & \textbf{11.517} & \textbf{3.932} & \underline{4.073} &\change{3.760} & \textbf{20.685} & \underline{0.287} \\
			& \textbf{Proposed }    & \textbf{3.560} & \underline{0.969} & \underline{11.453} & \underline{3.915} & \textbf{4.110} & \change{\underline{3.761}} & \underline{20.949} & \textbf{0.286}\\
			\midrule
			
			\multirow{5}{*}{DN+DR} 
			
			& UniverSE++          & 1.730 & 0.879 & 4.219 & 2.291 & 3.070 & \change{3.635} & 36.290 & 0.603 \\
			& MP-SENet Up.   &  2.330 & 0.925 & 8.589 & 3.165 & 3.578 & \change{\underline{3.990}} & 20.399 & 0.391 \\
			& SEMamba       & 2.372 & 0.929 & 8.747 & 3.160 & \textbf{3.792} & \change{\textbf{4.027}} & 19.365 & 0.364\\
			& ZipEnhancer   & \underline{2.401} & \underline{0.935} & \underline{9.204} & \underline{3.196} & 3.683 & \change{3.981} & \underline{18.847} & \underline{0.359}\\
			& \textbf{Proposed }   & \textbf{2.444} & \textbf{0.936} & \textbf{9.660} & \textbf{3.263} & \underline{3.733} & \change{3.979} & \textbf{18.095} & \textbf{0.344}\\
			\midrule
			
			\multirow{5}{*}{DN+DR+BWE} 
			& UniverSE++            & 1.610 & 0.871 & 3.938 & 2.117 & 2.987 & \change{3.580} & 39.195 & 0.634\\
			& MP-SENet Up.   & 2.103 & 0.916 & 6.792 & 2.546 & 3.376 & \change{\underline{3.841}} & 30.106 & 0.490 \\
			& SEMamba       & 2.066 & 0.921 & 6.617 & 2.712 & \underline{3.581} & \change{\underline{3.841}} & 29.371 & 0.485\\
			& ZipEnhancer   & \underline{2.169} & \underline{0.926} & \underline{7.116} & \underline{2.815} & 3.538 & \change{\textbf{3.844}} & \underline{29.100} & \underline{0.476} \\
			& \textbf{Proposed }  & \textbf{2.206} & \textbf{0.928} & \textbf{7.407} & \textbf{2.831} & \textbf{3.598} & \change{3.825} & \textbf{28.413} & \textbf{0.462}\\
			\bottomrule
		\end{tabular}
	\end{table*}
	\begin{table}[htbp]
	\centering
	\caption{Comparison of parameters and computational cost}
	\label{tab:param_compute_comparison}
	\resizebox{\columnwidth}{!}{
		\begin{tabular}{lccccc}
			\toprule
			\textbf{Metric} & \textbf{ZipEnhancer} & \textbf{UniverSE++} & \textbf{MP-SENet Up.} & \textbf{SEMamba} & \textbf{Proposed} \\
			\midrule
			\textbf{Params} (M) & 2.04 & 42.80 & 2.26 & 2.85 & \textbf{1.55} \\
			\textbf{MACs} (G/sec)  & \textbf{31.42} & 42.81 & 43.14 & 32.73 & {34.97} \\
			\bottomrule
		\end{tabular}
	}
	\end{table}
	In single-task evaluations, our method demonstrates \change{overall} superiority over strong baselines. For all tasks, our method outperforms the structural baseline MP-SENet Up across all \change{intrusive signal fidelity and phase accuracy metrics} with an evident margin.  Notably, UniverSE++ lags  behind the proposed method in signal fidelity metrics (e.g., PESQ and SI-SDR) and phase accuracy (PD). We attribute this to the inherent stochasticity of generative processes, which can introduce phase inconsistencies that degrade waveform alignment and consequently impair perceptual quality. For DN, our method achieves the best result in PESQ and phase accuracy, achieving parity with ZipEnhancer while generating clearer harmonic structures as evidenced in the first row of Fig. \ref{fig:wang6}. In DR, it secures the best performance across \change{almost} all metrics \change{with the exception of the reference-free DNSMOS, while delivering} significantly higher phase accuracy. In BWE, our model gets the top-tier performance comparable to ZipEnhancer, and shows a stronger ability to generate high-frequency speech components according to the third row of Fig. \ref{fig:wang6}.

The advantages of our approach are even more pronounced in the composite distortion scenarios, where our method achieves the best results across \change{the vast majority of} metrics \change{particularly dominating in intrusive metrics and phase precision}. \change{We note that our default model trails behind baselines on the reference-free DNSMOS metric. As detailed later in our ablation studies (Table \ref{tab:ablation_channelconf}), we observe a clear divergence between strict waveform fidelity and this specific metric. While scaling our model capacity generally improves intrusive perceptual metrics, phase precision, SI-SDR, and UTMOS, DNSMOS exhibits different behavior. Specifically, the highest DNSMOS score occurs in our lowest-capacity variant and surpasses all compared baselines in Table \ref{tab:universal_result} by a margin of over 0.1 despite yielding evidently sub-optimal scores across all other metrics, a phenomenon of metric divergence that has also been reported in recent literature \cite{zhang2025lessons}. Consequently, we view the sub-optimal DNSMOS scores of our default model in Table \ref{tab:universal_result} as a natural consequence of this metric divergence, rather than an architectural limitation.}
While the absolute improvements in \change{other} metrics may appear incremental, we conducted a paired t-test between the scores of the proposed method and ZipEnhancer, the second-strongest baseline, across the composite tasks. The results confirm that the performance gains are statistically significant ($p < 0.01$) for all metrics except \change{DNSMOS and} STOI\footref{web}.
This superiority might stem from the fact that our deep magnitude-phase manifold separation and GRE-based phase modeling has the potential to map the hidden features in a better way. Conventional predictive methods may struggle to map inputs to unconstrained high-dimensional manifolds that become brittle when different degradation overlaps with each other. In contrast, the proposed network bypasses unrelated manifold regions and focus capacity on valid speech reconstruction efforts, leading to better overall performance with comparable costs and fewer parameters.

	\subsection{Ablation Studies}
	\label{subsection: ablation studies}
	To ensure a rigorous evaluation, we focus on the composite DN+DR+BWE task for ablation studies, which represents the most complex degradation scenario.
Firstly, as in the phase retrieval task (Section \ref{subsection:phase retrieval}), we also perform an ablation to validate the rationale behind our global rotation equivariant design in composite degradation case.  
As shown in Table~\ref{tab:ablation_breakingRE}, 
\change{all modifications that violate GRE lead to degradations in phase-related metrics (PD and WOPD), confirming that rotation-equivariant constraints serve as an effective inductive bias for phase modeling. Breaking the FFN's equivariance yields a marginal improvement in UTMOS, but this comes at the cost of lower PESQ, STOI, PD, and WOPD. Because UTMOS is a reference-free model that primarily assesses overall naturalness and may be relatively insensitive to fine-grained phase difference, we interpret this isolated gain as 
	likely within the metric's inherent variance. Overall, the results consistently show that enforcing GRE across the network benefits phase estimation accuracy and generally lead to better speech quality.}
	\begin{table}[h]
		\centering
		\caption{Impact of violating Global Rotation Equivariance (GRE) in individual network components on model performance.}
		\begin{threeparttable}
			\resizebox{\columnwidth}{!}{%
				\renewcommand{\arraystretch}{1.3} 
				\begin{tabular}{l|ccccc}
					\hline
					\rule{0pt}{4ex}\textbf{Model} &  \textbf{PESQ} & \textbf{STOI}&  \textbf{UTMOS} & \textbf{PD($\downarrow$)} &
					\textbf{WOPD($\downarrow$)} \\[2ex] \hline		
					Proposed & 2.206 & 0.928 & 3.598 & 28.413 & 0.462 \\ 
					\hspace{0.3em}-- break MPICM GRE & 2.049  & 0.917 & 3.395 & 29.711 & 0.493 \\ 
					\hspace{0.3em}-- break Attn. GRE & \change{2.043} & \change{0.919} & \change{3.564} & \change{30.178} & \change{0.496}  \\ 
					\hspace{0.3em}-- break FFN GRE & \change{2.145} & \change{0.926} & \change{3.616} & \change{28.652}& \change{0.469}  \\ 
					\hline
				\end{tabular}%
			}
		\end{threeparttable}
		\label{tab:ablation_breakingRE}
	\end{table}
	
	We further investigate the impact of different convolutional channel configurations (Table \ref{tab:ablation_channelconf}). As expected, increasing channel depth generally improves performance; \change{Interestingly, the reference-free DNSMOS metric exhibits a diverging trend, with the smallest 32/16 configuration achieving the highest DNSMOS score (3.973) despite yielding the lowest scores across other metrics. We hypothesize that this phenomenon occurs because DNSMOS occasionally favors lower-capacity outputs that apply aggressively alter the spectrum to optimize for certain subjective perceptual patterns, even when these outputs are not fully aligned with the clean ground truth. Given this divergence, we selected the 48/16 configuration as the default model, as it secures evident gains in verifiable structural fidelity and phase accuracy while maintaining computational efficiency.} Notably, even our largest configuration maintains a lower computational cost than MP-SENet \change{Update}.
	While reducing the capacity of either stream leads to degradation \change{except DNSMOS}, shrinking the magnitude stream (e.g., 32/24) has a more pronounced impact than reducing the phase stream (48/16). This aligns with the primary role magnitude plays in speech perception. However, by isolating the circular phase topology, our architecture effectively removes phase-induced interference, enhancing magnitude modeling as evidenced by superior magnitude-dominated STOI scores reported before.
	
	\begin{table}[h]
		\centering
		\caption{An investigation of different channel configurations.}
		\begin{threeparttable}
			\footnotesize 
			\renewcommand{\arraystretch}{1.3} 
			\setlength{\tabcolsep}{4pt} 
			\resizebox{\columnwidth}{!}{%
			\begin{tabular}{l|cc|ccccccc}
				\hline
				\rule{0pt}{4ex}{\makecell{\textbf{Channels}\\ ($C_{\text{mag}}$/$C_{\text{pha}}$)}} & \textbf{\makecell{Para.\\ (M)}} & \textbf{\makecell{MACs\\ (G/sec)}} & \textbf{PESQ} & \textbf{STOI}& \change{\textbf{\makecell{SI-SDR\\(dB)}}} & \change{\makecell{\textbf{UT-}\\\textbf{MOS}}}& \change{\textbf{\makecell{DNS-\\MOS}}} & \textbf{\makecell{PD\\ ($\downarrow$)}}& \change{\textbf{\makecell{WOPD \\ ($\downarrow$)}}} \\[2ex] \hline		
				48 / 16 & 1.55  & 34.97 & \textbf{2.206} & \underline{0.928} & \change{\underline{7.407}} & \change{\underline{3.598}} & \change{3.825} & \underline{28.413} & \underline{0.462} \\ 
				32 / 16 & \textbf{0.90}  & \textbf{22.89} & 2.038 & 0.917 & \change{6.211}& \change{3.434} & \change{\textbf{3.973}} & 30.183 & \change{0.492} \\ 
				32 / 24 & 1.14 & 31.39 & 2.091 & 0.925 & \change{6.989} & \change{3.558} & \change{3.860} & 29.117 & \change{0.472}  \\ 
				48 / 24 & 1.78 & 42.72 & \change{\underline{2.164}} & \change{\textbf{0.933}} & \change{\textbf{7.719}} & \change{\textbf{3.687}} & \change{\underline{3.918}} & \change{\textbf{27.574}} & \change{\textbf{0.451}}  \\ 
				\hline
			\end{tabular}
		}
		\end{threeparttable}
		\label{tab:ablation_channelconf}
	\end{table}
	Next, we investigate the specific internal design choices of the MPICM. First, we validate the efficacy of the interactive gating mechanism by removing the cross-stream interaction entirely. As indicated in Table \ref{tab:ablation_mpicm_hadf}, this ablation results in a noticeable performance drop, confirming that the explicit exchange of information enables the two streams to mutually refine their latent magnitude and phase representations. 
	We also analyze the impact of the normalization strategy. While Complex RMS Norm is adopted for the phase branch to maintain GRE, we tested replacing the RMS Normalization in the magnitude branch with standard Instance Normalization. This change resulted in a performance decline. The  rationale of this phenomenon may lie in the non-negative magnitude input. Instance Normalization subtracts the mean, making the distribution center at zero in the very first layer, and may cause information loss when activated by SiLU. In contrast, RMS Normalization re-scales the energy and only alters the sign via bias, making it more stable for preserving the integrity of magnitude information in our separated magnitude-phase stream setting.
	
	\begin{table}[h]
		\centering
		\caption{Component-wise ablation study of the MPICM and HADF.}
		\begin{threeparttable}
			\resizebox{\columnwidth}{!}{%
				\renewcommand{\arraystretch}{1.3} 
				\begin{tabular}{l|ccccc}
					\hline
					\rule{0pt}{4ex}\textbf{Model} &  \textbf{PESQ} & \textbf{STOI}&  \textbf{UTMOS} & \textbf{PD($\downarrow$)} &
					\textbf{WOPD($\downarrow$)} \\[2ex] \hline		
					Proposed & 2.206 & 0.928 & 3.598 & 28.413 & 0.462  \\ 
					\hspace{1em}--w/o gating & 2.097 & 0.918 & 3.495 & 29.817& 0.490  \\ 
					\hspace{1em}-- MS$^*$ InstanceNorm & 2.069 & 0.920 & 3.330 & 29.149 & 0.481  \\ 
					\hspace{1em}-- w/o PS$^*$ Freq-wise Norm & 2.105 & 0.925 & 3.531 & 28.725 & 0.470 \\ \hline
					\hspace{1em}-- Naive Transformer-GRU & 1.995 & 0.909 & 3.126 & 31.236& 0.512  \\ 
					\hspace{1em}-- Merge Attention & 2.093 & 0.924 & 3.445 & 29.153 & 0.472  \\ 
					\hspace{1em}-- GRU-based Comp. FFN & 2.049 & 0.918 & 3.295 & 29.646 & 0.485  \\ 
					\hline
				\end{tabular}%
			}
			\begin{tablenotes}
				\footnotesize
				\item [$*$] MS/PS: Magnitude/Phase Stream.
			\end{tablenotes}
		\end{threeparttable}
		\label{tab:ablation_mpicm_hadf}
	\end{table}

	Turning to the bottleneck architecture, we first evaluate the necessity of our specialized dual-stream design by replacing the entire HADF module with the original, unified bottleneck from MP-SENet \change{Update}. As shown in Table \ref{tab:ablation_mpicm_hadf}, this reversion causes a significant performance degradation. This confirms that simply relying on general block to model the magnitude-phase dependencies is insufficient for dual-stream encoder-decoder.
	Next, we investigated the attention mechanism by evaluating a `Merged Attention' strategy. In this configuration, complex-valued phase features are mapped to the real domain via the modulus operation, concatenated with the magnitude features, and processed through a shared standard attention layer before branching into separate FFNs where the phase FFN uses the learned attention map to gate the complex features. The evident decline in all metrics indicates that collapsing the complex phase representation into a generic real-valued feature vector results in information loss, and the phase stream benefits from strictly preserving its complex-valued nature. This validates our Hybrid Attention design, which effectively unifies the confidence of both streams without compromising the geometric integrity. Finally, we evaluate the design of phase FFN by replacing the convolution pipeline with a GRU-based structure, analogous to that used in the magnitude branch. To mitigate the implementation complexity of a fully complex-valued GRU, we employed a real-valued GRU approximation operating on concatenated real and imaginary components. Empirical results demonstrate that our proposed convolutional phase FFN outperforms this recurrent alternative. This suggests that the phase stream benefits from the locality inherent in  convolution operations.

	\subsection{Visualization of the Attention Map}
\begin{figure}[t]
		\centering
		\includegraphics[width=\columnwidth]{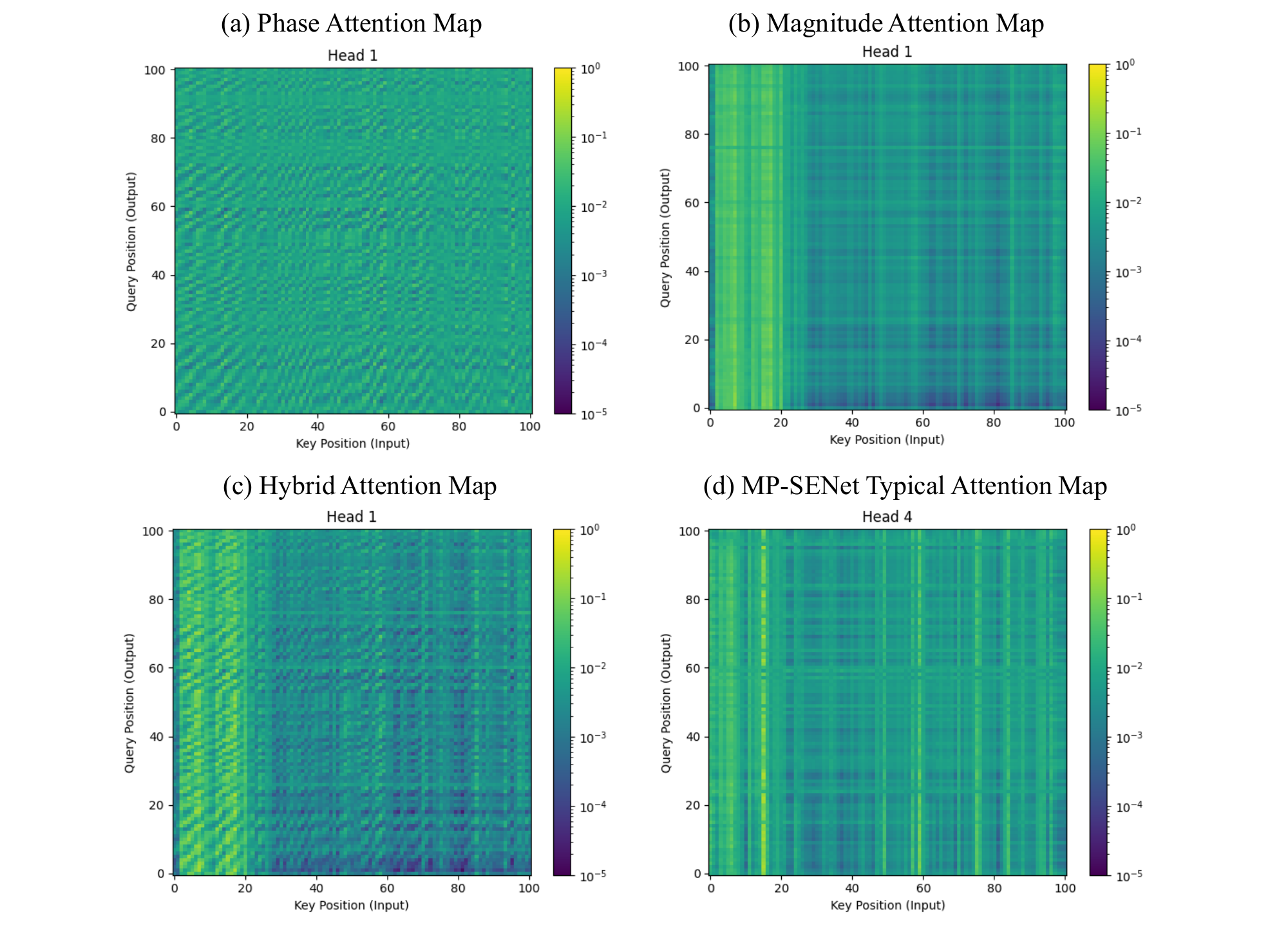}
		\caption{Visualization of learned attention patterns for a voiced speech segment. (\change{a}) The attention component derived solely from the phase stream. (\change{b}) The attention component derived from the magnitude stream. (\change{c}) The proposed Hybrid Attention map (fused). (\change{d}) The most correlated attention head from the MP-SENet \change{Update} baseline.}
		\label{fig:wang7}
	\end{figure}

	Finally, we visualize the attention maps from the HADF to analyze the internal mechanism of information fusion. We selected a representative frame containing a sustained voiced segment to highlight clear harmonic structures and plot its 2nd frequency-HADF attention map. By explicitly extracting the learned patterns from the magnitude and phase projections, distinct characteristics emerge, as shown in Fig.~{\ref{fig:wang7}}: the magnitude attention exhibits a \change{grid-shaped} distribution, capturing harmonic correlations and energy-based spectral dependencies across frequency bins. In contrast, the phase attention reveals a distinct periodic pattern, reflecting the angular similarities inherent to the circular manifold. For comparison, we analyzed the attention state of the same frame in MP-SENet \change{Update}, identifying the attention head with the closest pattern alignment to our learned features. Remarkably, its attention map shares a high degree of structural similarity with our magnitude pattern, indicating that joint mag-phase attention primarily captures energy-driven dependencies. However, the deviation of our composite Hybrid Attention score from the baseline map suggests that our method actively injects phase symmetry into the energy attention structure, which effectively allows the network to synthesize a dual-perspective representation.
	
	\section{Conclusion}
	In this paper, we have proposed \change{a speech enhancement method} that fundamentally rethinks phase modeling through the lens of its intrinsic circular topology, introducing a Global Rotation Equivariant architectural design. \change{By maintaining a magnitude-phase dual-stream structure and strictly enforcing geometric constraints on} the MPICM and HADF modules, our approach effectively eliminates the coordinate-system dependency for phase modeling inherent in conventional networks. Extensive evaluations across phase retrieval, denoising, and universal speech restoration tasks demonstrate that our method achieves \change{superior} performance with \change{state-of-the-art} phase reconstruction accuracy and generalization capability, all while maintaining comparable computational efficiency. Thorough ablations confirm that global rotation equivariance is an effective inductive bias to learn the intrinsic circular manifold of phase
\change{Future work will explore the extension of the proposed framework to multi-channel speech enhancement scenarios, alongside the development of lightweight, causal variants for real-time deployment.}
	
	
	\bibliographystyle{IEEEtran}
	\bibliography{reference}
\end{document}